\documentclass[a4]{article}

\usepackage[top=3cm, bottom=2cm, left=4cm, right=4cm]{geometry}

\usepackage[latin1]{inputenc}
\usepackage{amsfonts}
\usepackage{amsmath}
\usepackage{amssymb}
\usepackage{amsthm}
\usepackage[T1]{fontenc}
\usepackage{graphicx}

\theoremstyle{definition} 
\theoremstyle{definition} \newtheorem{defin}{Definition}
\theoremstyle{plain} \newtheorem{thm}{Theorem}[section]
\theoremstyle{plain} \newtheorem{lem}{Lemma}[section]
\theoremstyle{plain} \newtheorem{cor}{Corollary}[section]
\theoremstyle{plain} \newtheorem{prop}{Property}[section]
\theoremstyle{plain} \newtheorem{conj}{Conjecture}

\title{\sc{Random Walks, Electric Networks and The Transience Class problem of Sandpile\footnote{Extended draft appears in \cite{CS11}}}}
\author{Ayush Choure \quad Sundar Vishwanathan \\ Department of Computer Science and Engineering  \\ Indian Institute of Technology, Bombay \\ ayush, sundar@cse.iitb.ac.in}

\begin{document}
\maketitle

\begin{abstract}

The Abelian Sandpile Model is a discrete diffusion process defined on graphs (Dhar \cite{DD90}, Dhar et al. \cite{DD95}) which serves as the standard model of \textit{self-organized criticality}. The transience class of a sandpile is defined as the maximum number of particles that can be added without making the system recurrent (\cite{BT05}). We develop the theory of discrete diffusions in contrast to continuous harmonic functions on graphs and establish connections between standard results in the study of random walks on graphs and sandpiles on graphs. Using this connection and building other necessary machinery we improve the main result of Babai and Gorodezky (SODA 2007,\cite{LB07}) of the bound on the transience class of an $n \times n$ grid, from $O(n^{30})$ to $O(n^{7})$. Proving that the transience class is small validates the general notion that for most natural phenomenon, the time during which the system is transient is small. For degree bounded graphs, we demonstrate the first constant factor approximation algorithm for the transience class problem, based on harmonic functions. In addition, we use the machinery developed to prove a number of auxiliary results. We give general upper bounds on the transience class as a function of the number of edges to the sink. We exhibit an equivalence between two other tessellations of plane, the honeycomb and triangular lattices.

Further, for planar sandpiles we derive an explicit algebraic expression which provably approximates the transience class of $G$ to within $O(|E(G)|)$. This expression is based on the spectrum of the Laplacian of the dual of the graph $G$. We also show a lower bound of $\Omega(n^{3})$ on the transience class on the grid improving the obvious bound of $\Omega(n^{2})$.
\end{abstract} 

\section{Introduction}

The abelian sandpile model (ASM) is a type of diffusion process defined on graphs which is closely related  to  the  \textit{chip  firing game}  investigated  by  Bjorner, Lovasz and Shor \cite{auto} and Tardos \cite{chip}. Indeed some of the results in the model, proved by Biggs \cite{bigg}, have an analog in the ASM of Dhar \cite{DD90}. The  model proposed by Dhar has been studied in depth by the statistical physics community for investigating the phenomena known as \textit{self-organized criticality} in the dynamics of sandpile formation. This formulation is known to be closely related to other interesting albeit diverse phenomena such as stress distribution in earthquakes, size distribution in raindrops, path length distributions in loop-erased random walks, for instance. For a nice overview, see the recent comprehensive survey article by Dhar \cite{DD06}. The ASM, though easy to define, has a  very profound behavior and is far from being completely understood. Research in this area stretches across  numerous  disciplines such  as probability theory, algorithmics, theory of computing, combinatorics, non-linear dynamics, fractals, cellular automata, to name a few. These connections have been beautifully summarized by Kleber in \cite{gold}. Dhar \cite{DD06} also discusses some generalizations  of the ASM  like the Abelian  Distributed Processors (ADP) model which  is used to model a grid  of abstract state machines along with many theoretical and practical applications.

In the standard sandpile model, ``sand particles'' are added at the vertices of a (multi)graph. A site (vertex) is stable as long as the number of particles at the site remains less than its degree. Adding more particles would render the site unstable and is accompanied by the unstable site's passing a particle along each edge to its neighboring sites. This relaxation process is referred to as {\em{toppling}}. One of the sites known as the \textit{sink} cannot topple. To ensure that every relaxation process eventually stabilizes, one needs the condition that the sink is reachable from every other site. As the system evolves, the sandpile goes through a a sequence of configurations. Those which can be revisited in any toppling sequence are called {\em{recurrent}}, the remaining ones are termed {\em{transient}}. Typically, one starts with the empty configuration and as particles are added,one moves through transient configurations till a recurrent configuration is reached. Thereafter the configurations stay recurrent. The steady state behavior of a sandpile is characterized by its set of recurrent states. It has been observed by physicists that for most natural phenomena, the time taken to reach a recurrent state is small. Hence any acceptable model must reflect this tendency to reach steady state rapidly and it becomes important to study the time taken to reach recurrence in these models.

% behavior and an important parameter of study for these systems is the time taken to reach recurrence.
%For any acceptable model of natural phenomena, the time taken to reach a recurrent state can not to be too much.
% There is another important reason for studying the maximum time one can stay transient. The recurrence of any configuration is a combinatorial property, which follows from the efficient algorithm given by Dhar \cite{DD90} for deciding if the configuration is recurrent, now known as the \textit{burning test}.
% Recurrence can alternatively be defined in a more potential theoretic flavor. A configuration is recurrent if it can be obtained from any configuration via some toppling sequence which involves every site toppling at least once. The number of particles that need to be added at a particular site to observe a toppling at some other site is essentially related to the connectivity between two sites. We will discuss this intuitive connection and the formalizing arguments in greater details in coming sections.

The essential parameter in our discussion is the number of particles which ensure recurrence. If particles are added randomly then a simple coupon collector type argument demonstrates polynomial bounds on the \textit{expected} time to recurrence (as already mentioned in \cite{LB07}). The other scenario is to add particles adversatively so as to avoid a recurrent state for as long as possible. This problem was highlighted by Babai and Toumpakari \cite{BT05} where they define the requisite number of particles as the \textit{transience class} of the sandpile. This later motivated the insightful work by Babai and Gorodezky \cite{LB07} on grid based sandpile which are the most studied objects as compared to any other graph class because of their outstanding significance in statistical physics. In their path-breaking paper, Babai and Gorodezky \cite{LB07} show that for the standard $n \times n$ square grid based sandpile, the maximum number of particles one can add before hitting a recurrent state is $O(n^{30})$. This is a remarkable result in view of the fact that some closely related sandpile (for example line graph based) have transient state paths of length exponential in graph size. They use intricate combinatorial arguments based on particle conservation and the symmetry group of grid graphs to demonstrate the above mentioned bounds.
% along with other related conjectures and properties whose importance is both obvious and presented sufficiently well in the original manuscript.
However, simulations suggest a bound close to $O(n^{4})$ for the grid sandpile. Also the questions raised require analysis of the problem in a more general setting.

%a deeper
%It is this issue we address in this paper.
%It is hence important to know what the maximum time to reach recurrence is.

\textbf{Our contribution:} We begin by showing a strong connection between the transience class problem of sandpile and random walks on the underlying graph. Using LP duality and basic relaxation properties, we derive bounds on the transience class of a sandpile in terms of harmonic functions over the underlying graph. Similarly, bounds on sandpile impedances across any pairs of sites are obtained. These results form the core of our arguments which contrast the discrete model, sandpile, with the continuous version; random walks. For degree bound sandpile, we use the independent set properties of nodes with zero heights to demonstrate an algorithm which approximates the transience class up to constant factors. The algorithm works by computing harmonic functions over graphs and significantly tightens the connection between sandpile and random walks on the underlying graph. We then prove some basic properties of sandpile analogous to basic results in harmonic function theory, for example occurrence of worst case behavior at the boundary, reciprocity properties among any pair of sites, and so on. We derive and use a \textit{triangle inequality} of potentials. This inequality provides sufficient flexibility in analyzing the growth rates of harmonic functions at the cost of loosening the bounds. We use it to obtain a bound on the corner to corner potential response on a grid network. Using some symmetry property of grids, we prove the main result of our paper which improves the bound on transience class from $O(n^{30})$ (by Babai and Gorodezky \cite{LB07}) to $O(n^{7})$. We demonstrate a very general bound on the transience class in terms of sandpile size and the number of connections to the sink. We also show that in the case of planar sandpile, there exist explicit algebraic expressions which bound the transience class values. These are based on the spectrum of the Laplacian of planar dual of the given sandpile  graph. We derive the expression for grid sandpile and leave it as a (somewhat technical) conjecture to establish bounds on its value. We believe that these would yield bounds as low as $O(n^{4})$ for this problem. In the last section we discuss some important and interesting open problems that would be of interest to the theory community. Our main contribution in this paper is to bridge the gap between discrete diffusions on graphs and the theory of harmonic functions on graphs. Indeed, random walks, electric networks, graph spectra and LP duality have been central tools in theoretical computer science. We hope that this paper initiates a theory of discrete diffusions analogous to the celebrated theory of mixing of Markov chains.

%We also introduce the notion of \textit{transience class equivalence}, on the lines of complexity classes of algorithms and various categorizations of graph sequences. We demonstrate the power of our reduction by reducing a family of sandpiles into another, such that if transience class of one is polynomially bound, then so it is for the other. We conclude this discussion with an example of the other two regular tessellations of the plane, triangular and honeycomb lattices, by showing their equivalence.

\begin{subsection}{Related Work}
\textbf{Random walks and Sandpile:} We begin by sketching a picture depicting an intuitive connection between sandpile models and random walks on graphs. On a graph $G$ fix two vertices $s$ and $t$. A simple random walk with a specified starting vertex $v$ involves at each step, a choice of a neighboring vertex uniformly at random (See Bollob\'{a}s \cite{Bol} for a nice introduction). The {\em{potential}} $_{s}\pi_{t}(v)$ associated with $v$, and $s$ and $t$ as poles, is defined as the probability of reaching $t$ before $s$ starting from $v$. These $\pi$ functions, discussed at length in the next section are of paramount importance in analytic potential theory (see, for instance \cite{AT}). With site $s$ as the designated sink, add particles at site $v$ and observe the requisite number needed before a particle reaches site $t$. For any site which is ready to topple, if we label some particular particle among the set that are going to flow out, then the probability that it lands up at a particular neighbor, is uniform among the neighbors. If we add just enough particles at $v$, say $N_{v}$, so that exactly one particle reaches $t$, then the probability of it being any particular particle is uniform and the path it takes from $v$ to $t$ looks just like the ones constituting $_{s}\pi_{t}(v)$. Informally speaking, any particle in the starting pile at $v$ starts a random walk at $v$ which terminates at $t$ or $s$, whichever is encountered earlier. Intuitively, one expects that the probability $_{s}\pi_{t}(v)$ would be proportional to the reciprocal of $N_{v}$ with the proportionality factor accounting for discreteness and storage at sites, features absent in the usual network theory axioms. Our main theorem formalizes this connection and we derive as corollaries some properties of sandpile which are discrete analogues of the corresponding properties of random walks.\newline
\textbf{Electric Networks:} The classical theory of electric networks along with the well understood connections with random walks (\cite{NW}, \cite{DS},\cite{Lov}) has some very  powerful and intuitive results. These results have recently found applications in almost every important area of theoretical computer science. Christiano, Kelner, M\k{a}dry and Spielman \cite{CKMS} have recently announced the fastest known algorithm for computing approximate maximum $s-t$ flows in capacitated undirected graphs. Using the electric current flows in this network with $s$ and $t$ as poles, their algorithm constructs approximate flows. Earlier Kelner and M\k{a}dry \cite{KM} used arguments based on random walks to formulate the fastest known algorithm for generating spanning trees from uniform distribution. Spielman and Srivastava \cite{SS} construct good sparsifiers of weighted graphs via an efficient algorithm for computing approximate effective resistance between any two vertices, a result which is quite insightful on its own. The list of important results which use harmonic functions in an essential manner goes on. The benefit of this confluence of research in different classical areas is indeed mutual. For example, in their path breaking paper, Arora, Rao and Vazirani \cite{ARV} give an $O(\sqrt{\log n})$ approximation algorithm for computing graph conductance. Our goal has been to use the theory of harmonic functions in analyzing sandpile behavior (in the context of diffusion) in analogy with the theory of random walks on graphs. The results we report in this paper open up the possibility of analyzing those properties of ASM which may not have been possible using purely combinatorial arguments.   \newline
\textbf{Other Results on Sandpile:} As already mentioned, research problems on the abelian sandpile model span across numerous areas. Recent advances with a complexity theoretic flavor include proof of the one-dimensional sandpile prediction problem in \textbf{LOGDCFL} by Peter Bro Milterson \cite{Pet}. Also, Schulz \cite{MS} mentions a related NP-complete problem. The group structure of the space of recurrent configurations, first introduced by Dhar, Ruelle, Sen and Verma in \cite{DD95}, is also a fertile area of analysis. Cori and Rossin \cite{CR} show that sandpile groups of dual planar graphs are isomorphic. Toumpakari \cite{Tou} discusses some interesting properties of sandpile groups of regular trees where questions related to group rank are studied and the paper is concluded with an interesting conjecture on the rank of all Sylow subgroups of the sandpile group. Specific families of graphs like square cycles $C^{2}_{n}$, $K_{3} \times C_{n}$, $3 \times n$ twisted bracelets, etc have been analyzed. We refer the reader to \cite{CHW}, \cite{SH}, \cite{HL}.

%  We refer the reader the to \cite{CR}, \cite{Tou}, \cite{CHW}, \cite{SH} and \cite{HL}.
% Cori and Rossin \cite{CR} show that sandpile groups of dual planar graphs are isomorphic. Toumpakari \cite{Tou} discusses some interesting properties of sandpile groups of regular trees. Questions related to group rank are studied in particular, the paper is concluded with an interesting conjecture on the rank of all Sylow subgroups of the sandpile group. Specific families of graphs like square cycles $C^{2}_{n}$, $K_{3} \times C_{n}$, $3 \times n$ twisted bracelets, etc have been analyzed. We refer the reader to \cite{CHW}, \cite{SH}, \cite{HL}.
\end{subsection}

\section{Preliminaries}

\begin{subsection}{Introduction to the Abelian Sandpile Model}

Our notation and terminology follows Babai and Gorodezky \cite{LB07}.

\begin{defin}A {\em{graph}} $G$ is an ordered pair $(V(G),E(G))$ where $V(G)$ is called the set of vertices and $E(G)$ is a set of $2-$subsets of $V$, possibly with repeated elements, the set of edges.\end{defin}

This is referred to as a multi-graph in literature but we will use graph for brevity. The \textit{degree} of a vertex $v\in V$ is defined as the number of edges in $E$ which contain $v$.  Two vertices $v$ and $u$ are called \textit{adjacent} (or neighboring) if $(u,v) \in E$. A path between two vertices $u$ and $v$ is an ordered sequence of edges $e_{1}, e_{2}, \ldots, e_{k}$ such that $u \in e_{1}$, $v \in e_{k}$ and for all values of $i$, $e_{i} \cap e_{i+1} \neq \phi$. The graph $G$ is \textit{connected} if there exists a path between any pair of vertices.

To model an Abelian Sandpile Model, we take a connected graph $G$ with a special vertex called the \textit{sink}, denoted $s \in V$. Non-sink vertices in $G$ are called {\em{ordinary}} vertices and this subset will be denoted by $V_{o} = V - \{s\}$.

\begin{defin}The {\em{configuration}} of a sandpile $G$ is a map $c:V_{o} \rightarrow \mathbb{N}$, which will be represented as a vector. The weight of $c$ is $|c| = \sum_{v \in V_{o}}c(v)$. \end{defin}

The configuration $c$ records the number of sand particles contained in each of the ordinary sites. The \textit{empty} configuration is the zero vector. The \textit{capacity} of a site is the maximum number of particles that it can hold and is one less then the degree of the node.

\begin{defin}An ordinary node $v$ is said to be {\em{unstable}} in a configuration $c$ if $c(v) \geq \mbox{degree(v)}$. The configuration $c$ is said to be unstable if any site under it is unstable, else it is referred to as stable. \end{defin}

When a site is unstable it is said to \textit{topple}, that is it passes on some of its particles to its neighbors. When a site $v$ topples once, it loses $\mbox{degree(v)}$ particles and each neighbor of $v$ acquires a particle for every edge common with $v$. The sink node never topples. Starting with the empty configuration, we keep adding particles one by one on sites of our choice and topple them when necessary.

The ASM evolves in time through two modes, particle addition at sites and relaxation of unstable sites via topplings. A toppling sequence is an ordered set of configurations where every configuration can be obtained from the previous one by toppling some unstable site in it. Note that the event of many sites becoming unstable simultaneously poses no complication since the order in which they are subsequently relaxed does not affect the final stable configuration that is obtained at the end of toppling sequence. Elementary proofs of such confluence properties can be found in the pioneering paper on ASMs by Dhar \cite{DD90}. See also Babai and Toumpakari \cite{BT05}.

\noindent \textbf{Notation:} We write $c_{1} \geq c_{2}$ if $\forall v, c_{1}(v) \geq c_{2}$ and $c_{1} \vdash c_{2}$ if there is a toppling sequence which takes $c_{1}$ to $c_{2}$. Finally we write, $c_{1} \rightarrow c_{2}$ if $\exists c_{3} \geq c_{1}$ such that $c_{3} \vdash c_{2}$. We say that a configuration $c_{2}$ is \textit{reachable} from $c_{1}$ if $c_{1} \rightarrow c_{2}$ and \textit{unreachable} otherwise. In other words, one can add particles to certain sites in $c_{1}$ so that there exists a toppling sequence leading to $c_{2}$. Note that reachability is transitive, i.e. $c_{1} \rightarrow c_{2}, c_{2} \rightarrow c_{3} \Rightarrow c_{1} \rightarrow c_{3}$.

\begin{thm}(\cite{DD90},\cite{auto}) Given any configuration $c$, there exists a unique stable configuration $\sigma(c)$ such that $c \vdash \sigma(c)$, independent of the chosen toppling sequence.
\end{thm}

\begin{prop}If $c \vdash \sigma(c)$, then $kc \vdash k\sigma(c)$ \end{prop}

Associated with every toppling sequence is the count on the number of times each site has toppled, the vector of \textit{toppling potentials}, also referred to as the \textit{score vector} in \cite{LB07}. These toppling potentials are very closely related to the electric potentials that develop at various nodes when power source-sink are appropriately applied, a connection which we will discuss in detail in the coming sections.

\begin{defin} Assuming $c_{1} \vdash c_{2}$, the toppling potential function $z^{c_{1}, c_{2}} : V_{0} \rightarrow \mathbb{N}$ is defined as $z^{c_{1}, c_{2}}(v) :$ the number of times $v$ toppled in a toppling sequence from $c_{1}$ to $c_{2}$. We denote $z^{c, \sigma(c)}$ by $z^{c}$.\end{defin}

This function is well defined as the number of times a particular site topples is independent of the toppling sequence chosen, already noted in \cite{LB07}. The proof employs the fact that the principal minor of a connected graph's combinatorial laplacian is of full rank.

A configuration is called {\em{recurrent}} if it is reachable from \textit{any} configuration. As already mentioned, we say that a configuration $c_{i}$ is reachable from a configuration $c_{j}$ if by adding some particles to $c_{j}$ (possibly at multiple sites) and subsequently relaxing it, we can obtain $c_{i}$. A configuration is {\em{transient}} if it is not recurrent. The set of recurrent configurations is therefore, closed under reachability.
%
% \begin{prop} If configuration $c_{1} \leq c_{2}$, then recurrence of $c_{1}$ implies that of $c_{2}$.\end{prop}

\begin{prop}\label{prop:rec}If $\exists c'$ such that there is a toppling seqquence from $c'$ to $c$ in which every site has toppled at least once, then $c$ is recurrent.\end{prop}

The proof follows from the fact that the existence of such a toppling sequence precludes the existence of forbidden sub-configurations and hence makes the configuration recurrent. For a complete discussion on forbidden sub-configurations and recurrence of configurations the reader is referred to \cite{DD90} and \cite{DD06}.

We analyze the process of adding one grain at a time to the sandpile and study its evolution. As in the standard theory of Markov chains, recurrence characterizes the long term (steady state) behavior of sandpiles. Our investigation is concerned with the maximum number of particles that can be added while staying transient. Following Babai and Gorodezky \cite{LB07}, for a sandpile $S$ we define,

\begin{defin}The {\em{transience class}} of $S$ denoted by tcl($S$), is defined as the maximum number of particles that can be added to $S$ before reaching a recurrent configuration.\end{defin}

In view of property \ref{prop:rec}, we can bound the transience class from above by \textit{the maximum number of particles that can be added before all the nodes have toppled at least once}. Showing that this bound is tight upto constant factors is also not very hard. We defer a fuller disussion of questions of this nature to our subsequent manuscript.

\end{subsection}

\begin{subsection}{Basics of Harmonic Functions and Potential Theory}

For a very nice introduction to harmonic functions on graphs, we refer the reader to the beautifully written paper by Benjamini and Lovasz \cite{BL03} and to Telcs \cite{AT} for a thorough view. We start with some important definitions and fundamental properties. Given a connected graph $G$ and a function $\pi:V(G) \rightarrow \mathbb{R}$, we say that $\pi$ is {\em{harmonic}} over $V_{h}$ if,

\begin{eqnarray}
 \frac{1}{degree(v)}\sum_{u \sim v}\pi(u) = \pi(v) \quad v \in V_{h}
\end{eqnarray}

The remaining vertices (lying in $V - V_{h}$) are called the ``poles'' of $\pi$. The set $V_{h}$ is also called the {\em{interior}} of $\pi$ with vertices adjacent to the set of poles referred to as the {\em{boundary}}. We see that the value of $\pi$ at any vertex in $V_{h}$ is the average of its value in the immediate neighborhood. In case of multi graphs, we take the appropriate weighted means, where the weights are the number of common edges. This leads us to the first basic property,

\begin{prop}\label{prop:potMinMax} Any non-constant harmonic function can assume its extreme values only at the set of poles.\end{prop}

It follows that every non-constant harmonic function has at least two poles, its maxima and minima. Such functions are completely determined by their values on these vertices. Formally speaking,

\begin{prop}\label{prop:harm:uni1} Uniqueness: If two functions harmonic on $V_{h}$ agree on the boundary, they agree everywhere in the interior.\end{prop}

More generally, we have the following property,

\begin{prop}\label{prop:harm:uni2} Given a set of poles, a harmonic function is uniquely determined modulo scaling and translation by a constant.\end{prop}

Properties \ref{prop:harm:uni1} and \ref{prop:harm:uni2} important as they allow one considerable freedom in constructing harmonic completions of functions defined over the boundary set. This problem is the discrete analogue of the classical boundary value problems in complex analysis. We will describe two important examples in which these function arise naturally,

\textit{Random Walks on Graphs:} Consider a graph $G$ and two special vertices $s$ and $t$. The potential associated with $v$, with $s$ and $t$ as poles, $_{s}\pi_{t}(v)$ is defined as the probability of reaching $t$ before $s$ starting from $v$. One can check that the function $\pi$ so defined is indeed harmonic on the set $V - \{ s,t\}$, with the maximum value of $1$ at the node $t$ and the minimum value $0$ at $s$. The generalization to the multi-pole situation is also straightforward.

\textit{Electric Networks:} Consider a resistive electric network (i.e. a circuit made up entirely of resistors). Let $_{s}\pi_{t}(v)$ be the potential that appears at node $v$ when unit potential is applied across $t$ and $s$. Using the equation of charge conservation (Kirchoff's node law), one can show that these potentials are harmonic on all nodes except $s$ and $t$.

The main implication here is that one can intuitively think of the electric network theory as an analysis of random walks of electrons on the underlying graphs. Consequently, results from network theory can be used to prove interesting facts in other related areas. As an example, consider the problem of constructing the harmonic completion of a function with given boundary values. All one needs to do is to take the corresponding circuit and apply potentials equal to the boundary values on the boundary points. The potentials that will appear on other nodes can be computed using basic linear algebra (the only non-trivial step involves inverting the combinatorial Laplacian of $G$) thus allowing construction of harmonic completions efficiently. We outline below three very basic and fundamental results of network analysis which will be needed in the following sections.

\begin{thm}\label{thm:sup}{\textbf{Superposition Principle:}}
The superposition principle states that for all linear systems, the net response at a given place and time caused by two or more stimuli is the sum of the responses which would have been caused by each stimulus individually.
\end{thm}

\begin{thm}\label{thm:com}{\textbf{Compensation Theorem:}}
If the impedance $Z$ of a branch in a network in which a current $I$ flows is changed by a finite amount $dZ$, then the change in the currents in all other branches of the network may be calculated by inserting a voltage source of $-IdZ$ into that branch with all other voltage sources replaced by their internal impedances.
\end{thm}

\begin{thm}\label{thm:rec}{\textbf{Reciprocity Theorem:}}
In its simplest form, the reciprocity theorem states that if an emf E in one branch of a reciprocal network produces a current I in another, then if the emf E is moved from the first to the second branch, it will cause the same current in the first branch, where the emf has been replaced by a short circuit. Any network composed of linear, bilateral elements (such as R, L and C) is reciprocal.
\end{thm}

The reciprocity theorem can be restated in terms of just potential sources and potential measurements using the notion of effective resistances between pairs of nodes. The effective resistance between a pair of nodes $u$ and $v$, $R_{eff}(u,v)$ is defined as the potential difference which develops between $u$ and $v$ if a unit current source is applied across $u$ and $v$.

\begin{lem}\label{lem:potRec}Potential Reciprocity Lemma : If taking $s$ and $t$ as poles with $\pi(s) = 0$ and $\pi(t) = 1$ induces a potential of $_{s}\pi_{t}(v)$ at node $v$ and interchanging the roles of $v$ and $t$ induces $_{s}\pi_{v}(t)$ at $t$ then,
\begin{eqnarray}
 R_{eff}(s,t)_{s}\pi_{t}(v) =  R_{eff}(s,v)_{s}\pi_{v}(t)
\end{eqnarray}\end{lem}

\textbf{Proof }: Consider the given network $G$ with the special node $s \in V(G)$. We refer to the corresponding modified network $G(\epsilon)$ obtained from $G$ by adding an edge with resistance $1/\epsilon$ between every node and $s$. In particular, $G(0) \equiv G$. Furthermore, we refer to an edge between $s$ and $u$ by $\tilde{su}$. We will be using the current source version of the reciprocity theorem. If applying a unit current source across $\tilde{st}$ results in a potential of $v$ across $\tilde{sv}$, then applying unit current source across $\tilde{sv}$ results in a potential of $v$ units across  $\tilde{st}$. The value of this potential $v$ can be expressed, using Ohm's law, as the ratio of current through the edge and the resistance of the $\epsilon-$edge between the particular node and sink. Since both potentials are equal in magnitude, we can say that on $G(\epsilon)$,

\begin{eqnarray}
 R_{eff}(s,t)_{s}\pi_{t}(v) =  R_{eff}(s,v)_{s}\pi_{v}(t) \nonumber
\end{eqnarray}

This follows from observing that applying a unit current source across $\tilde{sv}$ is equivalent to applying a voltage source of $R_{eff}(s,v)$ across $\tilde{sv}$. Because of linearity, it follows that a potential of $_{s}\pi_{v}(t)R_{eff}(s,v).$  appears at node $t$. Similarly so for the other configuration.

This equation holds for arbitrarily small values of $\epsilon$. Consequently it holds for graph $G(0)$. $\blacksquare$

In particular, when the effective resistances across $s$ and $t$ are the same as $s$ and $v$, we have $_{s}\pi_{t}(v)  =   _{s}\pi_{v}(t)$. In the following discussion, we will omit the left subscript ($s$) from  $_{s}\pi_{t}$ whenever it is clear from context. We say that \textit{a walk $P$ is an instance of $_{s}\pi_{t}$} if it starts at some vertex $v$, avoids $s$ and ends at $t$. The following lemma may already be known to experts. Since we could not find it in literature, we present it with a simple proof.

\begin{lem}\label{lem:potTri} A triangle inequality for potentials
\begin{eqnarray}
\pi_{i}(j).\pi_{j}(k) \leq \pi_{i}(k)
\end{eqnarray}
\end{lem}
\textbf{Proof }: Let $\mathcal{P}(k,j)$ be the set of all walks from $k$ to $j$ avoiding $s$. This set can be partitioned into two components, namely the walks passing through $i$ and the ones avoiding it, denoted by $\mathcal{P}_{i}(k,j)$ and $\mathcal{P}_{\bar{i}}(k,j)$ respectively. For any walk $P$ which is an instance of $\pi_{i}$, let the probability of occurrence be $\pi_{i}(P)$. Then by definition,
\begin{eqnarray}
 \pi_{j}(k) = \sum_{P \in \mathcal{P}_{i}(k,j)}\pi_{j}(P) + \sum_{P \in \mathcal{P}_{\bar{i}}(k,j)}\pi_{j}(P) \nonumber
\end{eqnarray}

Similarly,
\begin{eqnarray}
 \pi_{i}(j) = \sum_{P \in \mathcal{P}(j,i)}\pi_{i}(P) \nonumber
\end{eqnarray}

Using these two relations, we obtain

\begin{eqnarray} \label{equ:potTriIneq1}
\pi_{i}(j).\pi_{j}(k) = \sum_{P \in \mathcal{P}(j,i)}\pi_{i}(P).\sum_{P \in \mathcal{P}_{i}(k,j)}\pi_{j}(P) + \sum_{P \in \mathcal{P}(j,i)}\pi_{i}(P).\sum_{P \in \mathcal{P}_{\bar{i}}(k,j)}\pi_{j}(P)
\end{eqnarray}

Consider the first term on right side in equation (\ref{equ:potTriIneq1}). Being a probability measure, the value of $\sum_{P \in \mathcal{P}(j,i)}\pi_{i}(P)$ is bounded above by $1$. Every $s$-avoiding walk from $k$ to $j$ passing through $i$ can be decomposed into two components, a walk from $k$ to $i$ avoiding $j$ and a walk from $i$ to $j$. This implies, $\sum_{P \in \mathcal{P}_{i}(k,j)}\pi_{j}(P) = \sum_{P \in \mathcal{P}_{\bar{j}}(k,i)}\pi_{i}(P).\sum_{P \in \mathcal{P}(i,j)}\pi_{j}(P) \leq \sum_{P \in \mathcal{P}_{\bar{j}}(k,i)}\pi_{i}(P)$. The first term therefore has the following bounds,

\begin{eqnarray}\label{equ:potTriIneq2}
 \sum_{P \in \mathcal{P}(j,i)}\pi_{i}(P).\sum_{P \in \mathcal{P}_{i}(k,j)}\pi_{j}(P) \leq \sum_{P \in \mathcal{P}_{\bar{j}}(k,i)}\pi_{i}(P)
\end{eqnarray}

For bounding the second term, observe that any $s$-avoiding walk from $k$ to $i$ which passes through $j$, can be treated as a juxtaposition of a walk from $k$ to $j$, avoiding $i$, and a walk from $j$ to $i$. Hence,

\begin{eqnarray}\label{equ:potTriIneq3}
 \sum_{P \in \mathcal{P}(j,i)}\pi_{i}(P).\sum_{P \in \mathcal{P}_{\bar{i}}(k,j)}\pi_{j}(P) & = & \sum_{P \in \mathcal{P}_{j}(k,i)}\pi_{i}(P)
\end{eqnarray}

Using equations \ref{equ:potTriIneq1}, \ref{equ:potTriIneq2} and \ref{equ:potTriIneq3} we get

\begin{eqnarray} \label{equ:potTriIneq4}
\pi_{i}(j).\pi_{j}(k) \leq \sum_{P \in \mathcal{P}_{j}(k,i)}\pi_{i}(P) + \sum_{P \in \mathcal{P}_{\bar{j}}(k,i)}\pi_{i}(P) = \pi_{i}(k) \nonumber
\end{eqnarray}
$\blacksquare$ \newline

\noindent \textit{Remark:} The utility of this inequality becomes clear when interpreted in the context of electric networks. Consider a network such that the node with ground potential is fixed and we are allowed to apply power at any other node and observe the resulting potentials. The inequality implies that if applying a potential $V_{1}$ at $i$ produces unit potential at node $j$ and applying $V_{2}$ at node $j$ produces unit potential at node $k$, then applying $V_{1}.V_{2}$ units at $i$ produces \textit{at least} unit potential at node $k$.

\end{subsection}

\section{Reducing the transience class problem to estimating harmonic functions over graphs}

We first consider the single site particle addition strategies. We will later show that the effect of allowing particle addition at multiple sites on our transience class estimates is inconsequential as far as our estimates are concerned.

\begin{defin} Consider a sandpile $S$ with nodes $u$ and $w$. The {\em{sandpile impedance}} of the ordered pair $(v,w)$, $R_{s}(v,w)$ is defined as the maximum number of particles that one can add at $v$ before a toppling at $w$ occurs.\end{defin}

Note that unlike the impedance of electric networks, sandpile impedance is not symmetric in its arguments, i.e. in general $R_{s}(v,w) \neq R_{s}(w,v)$. To estimate its value, we introduce the following LP relaxation.

\begin{eqnarray}
\text{max } x_{v} \nonumber \\
  0 \leq \sum_{v' \sim v} z(v') - d(v).z(v)  + x_{v} & \leq & d(v) - 1\nonumber \\
  \forall u \neq v : 0 \leq \sum_{u' \sim u} z(u') - d(u).z(u)  & \leq & d(u) - 1\nonumber \\
  z(w) \leq 0, z \geq 0, x \geq 0 \nonumber
\end{eqnarray}

The values of $x_{v}$ (the number of particles added at $v$) and $z$ (the vector of toppling counts) that are realized above are a feasible solution of this LP and hence the optimum of this LP yields an upper bound on the $R_{s}(v,w)$. With the fixed sink node, $s$, we define $\pi_{w}(v)$ as the potential at node $v$ when a unit potential is applied at node $w$. In terms of these potential functions, the following bound holds.

\begin{lem}\label{lem:tclUpp} The optimum value of the above LP is bounded from above by the following value,
 \begin{eqnarray}\label{equ:tclUpp}
\frac{1}{\pi_{w}(v)} \sum_{u}(d(u)-1).\pi_{w}(u)
\end{eqnarray}
\end{lem}

\textbf{Proof:}  We consider the following relaxed version of the given LP.
\begin{eqnarray}
\text{max } x_{v} \nonumber \\
\sum_{v' \sim v} z(v') - d(v).z(v)  + x_{v} & \leq & d(v) - 1\nonumber \\
  \forall u \neq v :\sum_{u' \sim u} z(u') - d(u).z(u)  & \leq & d(u) - 1\nonumber \\
  z(w) \leq 0, z \geq 0, x \geq 0 \nonumber
\end{eqnarray}

From the weak duality for LPs, it follows that to obtain an upper bound of $\alpha$ on the optimum value of the above system, it suffices to find a feasible solution of the dual LP of value $\alpha$. The dual is the following :

\begin{eqnarray}
\text{min } \sum_{u}(d(u)-1).Y(u)  \nonumber \\
\sum_{u' \sim w} Y(u') + Y' - d(w).Y(w) & \geq & 0 \nonumber \\
\forall u \neq w : \sum_{u' \sim u} Y(u') - d(u).Y(u) & \geq & 0 \nonumber \\
 Y(v) \geq 1, Y \geq 0, Y' \geq 0  \nonumber
\end{eqnarray}

Consider the following set of equations

\begin{eqnarray}
\sum_{u' \sim w} Y(u') + Y' - d(w).Y(w) & = & 0  \label{equ:curInp} \\
\forall u \neq w : \sum_{u' \sim u} Y(u') - d(u).Y(u) & = & 0 \\
Y(v)  & = & 1 \nonumber
\end{eqnarray}

A non-negative set of values satisfying the above set is feasible for the dual LP. We find these by considering the resistive circuit $\widehat{S}$, obtained by replacing each edge in $S$ by a unit resistance. We assign ground potential to the sink, and inject current at node $w$ such that it gets unit potential. The potential that develops on any node $u$ is $_{s}\pi_{w}(u)$. The potential value at node $v$, $\pi_{w}(v)$, can be used to scale the input current at $w$ thereby scaling all the potentials as well, such that potential at node $v$ becomes unit. It follows that the values $Y(u) = \pi_{w}(u)/\pi_{w}(v)$ and $Y'$ equaling the value of the current injected form a feasible solution of the dual LP. The objective value at this point is,

\begin{eqnarray}
\frac{1}{\pi_{w}(v)} \sum_{u}(d(u)-1).\pi_{w}(u) \nonumber
\end{eqnarray}

$\blacksquare$

This yields an upper bound on $x_{v}$. To obtain a lower bound, consider the complementary problem of finding $x_{v}'$, the \textit{minimum} number of particles that must be added at $v$ to observe a toppling at $w$. The following LP's objective value forms a lower bound on $x_{v}'$,

\begin{eqnarray}
\text{min}\{ x_{v}' \} \nonumber \\
  0 \leq \sum_{v' \sim v} z(v') - d(v).z(v)  + x_{v}' & \leq & d(v) - 1\nonumber \\
  \forall u \neq v : 0 \leq \sum_{u' \sim u} z(u') - d(u).z(u)  & \leq & d(u) - 1\nonumber \\
  z(w) \geq 1, z \geq 0, x \geq 0  \nonumber
\end{eqnarray}

The proof of the following lemma is analogous to the previous case.

\begin{lem}\label{lem:tclBel} The optimum value of the above LP is bounded from below by the following value,
\begin{eqnarray}\label{equ:tclBel}
\frac{1}{\pi_{w}(v)}
\end{eqnarray}
\end{lem}

\textbf{Proof:} Consider the relaxed version of the above LP,

\begin{eqnarray}
  \text{min} \{ x_{v}' \} \nonumber \\
   \sum_{v' \sim v} z(v') - d(v).z(v)  + x_{v}' & \geq & 0 \nonumber \\
  \forall u \neq v :\sum_{u' \sim u} z(u') - d(u).z(u)   & \geq & 0 \nonumber \\
  z(w)  \geq  1, z \geq  0, x \geq  0 \nonumber
\end{eqnarray}

From the weak duality for LPs, it follows that to obtain a lower bound on the optimum value of the above system, it suffices to find a feasible solution of the dual LP. The dual is the following :

%\begin{eqnarray}
%\text{max } Y(v)  \nonumber \\
%\sum_{u' \sim w} Y(u') + Y' - d(w).Y(w) & \geq & 0 \nonumber \\
%\forall u \neq w : \sum_{u' \sim u} Y(u') - d(u).Y(u) & \geq & 0 \nonumber \\
% Y(v) \geq 1, Y \geq 0, Y' \geq 0  \nonumber
%\end{eqnarray}

\begin{eqnarray}
\text{max } Y(w)  \nonumber \\
\sum_{u' \sim w} Y(u') + Y' - d(w).Y(w) & \leq& 0 \nonumber \\
\forall u \neq w : \sum_{u' \sim u} Y(u') - d(u).Y(u) & \leq & 0 \nonumber \\
 Y(v) \leq 1, Y \geq 0, Y' \geq 0  \nonumber
\end{eqnarray}

Consider the following set of equations

\begin{eqnarray}
\sum_{u' \sim w} Y(u') + Y' - d(w).Y(w) & = & 0  \label{equ:curInp} \\
\forall u \neq w : \sum_{u' \sim u} Y(u') - d(u).Y(u) & = & 0 \\
Y(v)  & = & 1 \nonumber
\end{eqnarray}

As before, any non-negative set of values satisfying the above system is feasible for the dual LP, and therefore forms a lower boudn on the objective value. We find these by considering the resistive circuit $\widehat{S}$, obtained by replacing each edge in $S$ by a unit resistance. We assign ground potential to the sink, and inject current at node $w$ such that it gets unit potential. The potential that develops on any node $u$ is $_{s}\pi_{w}(u)$. The potential value at node $v$, $\pi_{w}(v)$, can be used to scale the input current at $w$ thereby scaling all the potentials as well, such that potential at node $v$ becomes unit. It follows that the values $Y(u) = \pi_{w}(u)/\pi_{w}(v)$ and $Y'$ equaling the value of the current injected form a feasible solution of the dual LP. The objective value at this point is $\pi_{w}(v)^{-1}$. $\blacksquare$

Clearly the maximum number of particles that can be added at $v$ before toppling some $w$ is just one less then the minimum number that need to be added at $v$ to topple $w$, that is $x_{v}' =  x_{v} + 1$. Using equations (\ref{equ:tclUpp}) and (\ref{equ:tclBel}), the following two-sided bounds are obtained.

\begin{eqnarray}
\frac{1}{\pi_{w}(v)} - 1 \leq x_{v} \leq \frac{1}{\pi_{w}(v)} \sum_{u}(d(u)-1).\pi_{w}(u)
\end{eqnarray}

Further, define the \textit{potential profile} of the circuit when unit potential is applied at node $w$ as
\begin{eqnarray}
  \Gamma_{S}(w) = \sum_{v}(d(v)-1).\pi_{w}(v) \nonumber
\end{eqnarray}

Using this notation, $R_{S}(v,w) = x_{v}$ satisfies the following general bounds.

\begin{lem}\label{lem:tclsi} $R_{s}(v,w)$ is $O(\Gamma_{S}(w).\pi_{w}(v)^{-1})$. \end{lem}

%When $v$ is a site adjacent to the sink, $d(v) - \sum_{v' \sim v} \pi_{v}(v')$ is at least unit ($v$ is at unit potential adjacent to the sink whose potential is zero).

\begin{lem}\label{lem:tclsil} $R_{s}(v,w)$ is $\Omega(\pi_{w}(v)^{-1})$. \end{lem} %, when $v$ is adjacent to the sink. \end{lem}

To find the maximum number of particles one can add at $v$ before every other site topples, one simply needs to consider the maximum value of $R_{S}(v,w)$ over all values of $w \in V_{o}$. Consequently, one can find the maximum number of particles that can be added at a \textit{single} site before every other site topples, by considering the maximum of $R_{S}(v,w)$ over all pairs $(v,w)$. This value, $\max_{v,w}\{\Gamma_{S}(w)\pi_{w}(v)^{-1}\}$, also forms a bound on $tcl(S)$ as allowing particle addition at multiple sites above gives the same estimates. This fact follows from essentially the same line of argument that was used for finding the upper bounds except that in this case instead of $x_{v}$, the objective function to maximize is $\sum_{u}x_{u}$ where $x_{u}$ is the number of particles added at $u$.

\begin{thm}\label{thm:tcl} $tcl(S)$ is $O(\max_{v,w}\{\Gamma_{w}(S)\pi_{w}(v)^{-1}\})$. \end{thm}

As each of the $\pi_{w}(v)$ lies between $0$ and $1$, the value of $\Gamma_{w}(S)$ is therefore bounded between $1$ and $|2E(S)|$. Hence, we have the following relaxed  upper bound on the value of $tcl(S)$,

\begin{lem}\label{lem:tcl} $tcl(S)$ is $O(|E(S)|.\max_{v,w}\{\pi_{w}(v)^{-1}\})$. \end{lem}

These results quantify the relationship between sandpiles and random walks on graphs. However, the theory of potential functions on graphs boasts of several very intuitive and beautiful results, e.g. the recipocity theorem. In a later section, we show that the parallelism between sandpiles and electric networks runs deeper by demonstrating the sandpile versions of some well known basic results in network theory.

We start by considering some simple properties of potential funtions. For example, consider the property \ref{prop:potMinMax} which says that the maximum and minimum of potential functions occur at poles. An elementary proof by contradiction is easily conceivable. In the case of sandpiles, one can think of a similar notion of maxima/minima in terms of ease of percolation of particles. When we add particles anywhere (may be more than one site), then is it so that the last site to topple will be adjacent to sink? Or consider the dual problem. We are allowed to add particles at one site only. For any particular site whose toppling we wish to delay for as long as possible, is it true that the best strategy is to add particles at a site adjacent to sink? To rephrase, is it true that for any site $w$, the value of $R_{S}(v,w)$ is maximized for some $v$ adjacent to the sink?

Note that both these questions are two sides of the same coin in case of random walks because of the reciprocity properties discussed above. The first has a direct analogy for sandpiles.

\begin{lem}\label{prop:boundSand1} The last site to topple is always adjacent to sink.\end{lem}
\textbf{Proof }: Assume that the last site to topple is not adjacent to sink. Since each of its neighbors has already toppled, it has received at least as many particles as its degree and has become unstable at least once contradicting the assumption that the particular site has never toppled. $\blacksquare$

We now fix the site under observation and ask the same question about the site where we add particles.

\begin{lem}\label{prop:boundSand2} For a given $w$, the estimate of $R_{s}(v,w)$ is maximum when $v$ is at boundary. \end{lem}
\textbf{Proof }:For a fixed $w$, the value of $\Gamma_{S}(w)$ is fixed. One has to show that the value of $\pi_{w}(v)$ is minimum for some vertex $v$ adjacent to sink, $s$. This clearly follows from the fact that for every internal node $u$, $\pi(u)$ is a convex combination, in particular the weighted arithmetic mean, of the $\pi(.)$ values at its neighbors. This means that $\pi(u)$ is bounded between the values spanned by the neighbors, so it cannot be an extreme point.
 $\blacksquare$

\noindent \textit{Note:} This lemma talks about the estimate and \textit{not} the exact value of $R_{s}(v,w)$. The lemmas prove that while using the Theorem \ref{thm:tcl}, it is enough to consider both sites on the boundary set (i.e. adjacent to sink). The following lemma is the sandpile analogue of the classical potential reciprocity lemma from network theory.

\begin{lem}Sandpile Reciprocity lemma : If adding $p$ particles at $v$ causes toppling at $w$ then adding $2|E(S)|\frac{R_{eff}(v,s)}{R_{eff}(w,s)}.p$ particles at $w$ causes a toppling at $v$.\end{lem}

\textbf{Proof }: Using theorem \ref{lem:tclsi}, the ratio of $R_{S}(v,w)$ to $R_{S}(w,v)$ can be bounded.

\begin{eqnarray}
 \frac{R_{S}(w,v)}{R_{S}(v,w)} \leq \frac{\max R_{S}(w,v)}{\min R_{S}(v,w)} = \frac{\pi_{v}(w)^{-1}.\Gamma(v)}{\pi_{w}(v)^{-1}} \leq 2|E(S)|.\frac{R_{eff}(v,s)}{R_{eff}(w,s)} \nonumber
\end{eqnarray}

Where the last inequality follows from the potential reciprocity mentioned in Lemma \ref{lem:potRec}. Given $p = R_{S}(v,w)$, we get the required bound on $R_{S}(w,v)$ in terms of $p$.

$\blacksquare$

One can go even further by using the fact that the maximum value of $R_{eff}$ is $|V(s)|$ (attained for paths) and minimum value is at least $1/|E(S)|$ (attained for just a pair of adjacent nodes with many parallel edes betwene them). Babai and Gorodezky \cite{LB07} conjectured the following.

\begin{conj}(\cite{LB07})Assume that for sandpile $\chi$, the induced sub-graph on the set of ordinary vertices is connected. Then the transience class of $\chi$ (the largest weight of any transient configuration) is the height of the tallest transient stack of grains placed on a single site.\end{conj}

The conjecture is equivalent to saying that using single site particle addition strategies one can attain the transience class bounds. However this is not so. A counter-example by Sunic appears in \cite{BT05}. We present a simpler counterexample and an intuitive reason why this conjecture is false in general. Assume that in a sandpile $\chi$ with finite transience class, one can attain the bounds by adding particles at the single site $v$ and $w$ topples last. After the last particle is added, adding one more is supposed to make the configuration recurrent. Which means this heaviest transient configuration, when relaxed should have every site filled to its maximum capacity, except for $w$ which has not yet toppled and contains particles less then the maximum capacity. If there exists some other site which is not filled up to maximum capacity, one can add particles there and fill it up (only till it stays stable, of course). Hence the validity of conjecture rests on the rather unlikely premise that in the heaviest transient configuration, every site but one is filled to its maximum stable capacity. A condition which one would think unlikely when there are no symmetries in $\chi$ (trivial automorphism groups). We will present an example with non-trivial symmetries to demonstrate that even in this case, one cannot expect such a strong property.

Consider the grid sandpile $\chi_{n}$ with $n = 4$. We add particles at the top left corner. As expected, the last site to topple is the bottom right corner. Here is the stable configuration corresponding to the heaviest transient configuration with a stack of particles placed on the top left.
\begin{eqnarray}
3 \phantom{a} 3 \phantom{a} 3 \phantom{a} 0 \nonumber \\
3 \phantom{a} 0 \phantom{a} 3 \phantom{a} 2 \nonumber \\
3 \phantom{a} 3 \phantom{a} 2 \phantom{a} 3 \nonumber \\
0 \phantom{a} 2 \phantom{a} 3 \phantom{a} 2 \nonumber
\end{eqnarray}

All the sites which have fewer then $3$ particles, can be topped up without inducing a toppling at the bottom right corner node, and then adding a particle at top left corner induces recurrence. This demonstrates that single site particle addition do not work for this pair. We discuss a related open question in the section of future work.

% We believe however that the following weaker conjecture holds.
%
% \begin{conj}\label{conj:tclSing} Assume that for sandpile $\chi$, the induced sub-graph on the set of ordinary vertices is connected. Then the transience class of $\chi$ (the largest weight of any transient configuration) is bounded from above by the sum of height of the tallest transient stack of grains placed on a single site and the size of graph $\chi$.\end{conj}
%
% \noindent \textit{Remark:} One can show that the transience class of a sandpile is bounded from below, up to constant factor, by the graph size. This follows from the fact that when every site has toppled, no two adjacent sites can both have zero particles. Which implies that for every edge, at least one particle is on board. So $tcl(\chi) = \Omega(|\chi|)$. In the light of this observation, conjecture \ref{conj:tclSing} means that the esimates derived using single site particle addition strategies are constant additive factor approximations of the actual transience class. Also note that the above conjecture is stronger then the question raised by Sunic, mentioned in \cite{BT05}, that local (i.e. single site particles addition type) transience classes are bounded by $0.5$-factor approximations of the transience class, as we have an additive error term compared to the previous multiplicative one.

\section{A constant factor approximation for transience classes of degree bounded graphs}

We will now argue that in the case of degree bound graphs, using some combinatorial properties of independent sets in sandpile graphs along with arguments similar to the ones outlined in the previous section for the lower bound derived in Lemma \ref{lem:tclBel}, one can derive tight (up to constant factors) lower bounds for the transience classes.

In the sandpile $S$, let $x_{v}'$ be the minimum number of particles that can need to be added at node $v$ such that the node $w$ topples at least once. Clearly $x_{v}' = x_{v} + 1$.  As in the previous section, the following LP relaxation forms a lower bound on the value of $x_{v}'$.

\begin{eqnarray}
\text{min } x_{v}' \nonumber \\
0 \leq \sum_{v' \sim v} z(v') - \text{degree}(v).z(v)  + x_{v} & \leq & \text{degree}(v) - 1\nonumber \\
  \forall u \neq v :0 \leq \sum_{u' \sim u} z(u') - \text{degree}(u).z(u)  & \leq & \text{degree}(u) - 1\nonumber \\
  z(w) \geq 1, z \geq 0, x_{v}' \geq 0 \nonumber
\end{eqnarray}

Consider the scenario in which we add $x_{v}'$ particles at $v$ and allow the configuration to settle down to stability. Furthermore, let the number of particles that appear at any node $u$ in the resulting stable configuration be $h_{u}$. The following lemma gives a lower bound on the particle count $x_{v}'$ in terms of these height functions $h_{u}$.

\begin{lem}\label{lem:tclLow} The value of the particle count $x_{v}'$ is bounded from below by the following value,
 \begin{eqnarray}\label{equ:tclLow}
\frac{1}{\pi_{w}(v)} \sum_{u} h(u).\pi_{w}(u)
\end{eqnarray}
\end{lem}

\noindent \textbf{Proof:} Clearly $0 \leq h_{u} \leq \text{degree}(u)$. In the linear program stated above, replacing each pair of constraints of the type  $0 \leq \sum_{v' \sim v} z(v') - d(v).z(v)  + x_{v}  \leq  d(v) - 1$ by $\sum_{v' \sim v} z(v') - d(v).z(v)  + x_{v} = h_{v} $, maintains the property that the optimum value is at least a lower bound to the \textit{exact} solution to the particle count $x_{v}'$. The altered LP is the following,

\begin{eqnarray}
\text{min } x_{v}' \nonumber \\
\sum_{v' \sim v} z(v') - \text{degree}(v).z(v)  + x_{v}' & = & h(v) \nonumber \\
  \forall u \neq v :\sum_{u' \sim u} z(u') - \text{degree}(u).z(u)  & = & h(u)\nonumber \\
  z(w) \geq 1, z \geq 0, x_{v}' \geq 0 \nonumber
\end{eqnarray}

To bound the optimum value, we will consider the dual of this minimization program and find a suitable feasible point. The value of the cost function at that point will be used as bound. The dual is the following maximization program.

\begin{eqnarray}
\text{max } Y' + \sum_{u} h(u).Y(u)  \nonumber \\
\sum_{u' \sim w} Y(u') + Y' - \text{degree}(w).Y(w) & \leq & 0 \nonumber \\
\forall u \neq w : \sum_{u' \sim u} Y(u') - \text{degree}(u).Y(u) & \leq & 0 \nonumber \\
 Y(v) \leq 1, Y \geq 0, Y' \geq 0  \nonumber
\end{eqnarray}

As in the previous lemma, we consider the following set of equations, whose feasible region is inside the one corresponding to the dual we mentioned above.

\begin{eqnarray}
\sum_{u' \sim w} Y(u') + Y' - d(w).Y(w) & = & 0  \nonumber \\
\forall u \neq w : \sum_{u' \sim u} Y(u') - \text{degree}(u).Y(u) & = & 0 \nonumber \\
Y(v)  & = & 1 \nonumber
\end{eqnarray}

A non-negative set of values satisfying the above set is feasible for the dual LP. We find these by considering the resistive circuit $\widehat{S}$ corresponding to the graph $S$ (with each edge having unit resistance). The sink node is assigned ground potential and just enough current is injected at node $w$ so that it attains unit potential. In terms of the standard potential functions described earlier, the potential that develops on any node $u$ is $_{s}\pi_{w}(u)$. The potential value at node $v$, $\pi_{w}(v)$, can be used to scale the input current at $w$ thereby scaling all the potentials as well, such that potential at node $v$ becomes unit. It follows that the values $Y(u) = \pi_{w}(u)/\pi_{w}(v)$ and $Y'$ equaling the value of the current injected form a feasible solution of the dual LP. The objective value at this point is,

\begin{eqnarray}
Y' + \frac{1}{\pi_{w}(v)} \sum_{u}h(u).\pi_{w}(u) \nonumber
\end{eqnarray}

Since the current, $Y'$, is a positive quantity, the above derivation implies the Lemma.$\square$

We will now show that the lower bound is at most a constant factor smaller then the upper bound. We start with defining an auxiliary set of variables $\hat{h}_{v}$ such that,

\begin{eqnarray}
\hat{h}_{v} = 1 \equiv h_{v} \geq 1 \nonumber
\end{eqnarray}

From the positivity of $\pi(.)$ function and the domination relation $\hat{h}_{v} \leq h_{v}$, the following inequality follows.

\begin{eqnarray}\label{equ:tclLower2}
\widehat{\Gamma} = \frac{1}{\pi_{w}(v)} \sum_{u}\hat{h}_{u}.\pi_{w}(u) \leq \frac{1}{\pi_{w}(v)} \sum_{u}h_{u}.\pi_{w}(u)
\end{eqnarray}

We will now bound the value $\widehat{\Gamma}$ from below.

\begin{lem}(Dhar \cite{DD06}, Babai and Gorodezky \cite{LB07})\label{lem:indSet} In any stable recurrent configuration, for every edge, both the incident vertices cannot have zero particles.\end{lem}

This follows from the fact that in any stable recurrent configuration, the last toppling of one of them would have taken place after the other and so the second node necessarily has at least one particle. For every edge $u,v$, at least one of $h_{u}$ and $h_{v}$ is $\geq 1$.

\begin{cor} The set of nodes $I = \{v : h_{v} = 0\}$ form an independent set.\end{cor}

\begin{figure}[ht]
\centering
\includegraphics[width=.5\textwidth]{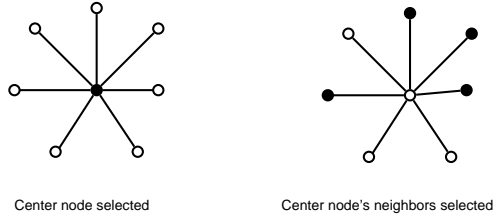}
\caption{Neighborhood around vertex $v$; two cases, $v$ selected and not selected in $I$}
\label{fig:localNei}
\end{figure}

Assume further that the graph satisfies ($\Delta$), i.e. the maximum degree is $\Delta$. Using this we will obtain the required bounds. Consider any vertex $v \in V_{h}$ along with its neighborhood (see figure (\ref{fig:localNei})). The function $\pi(.)$ is harmonic over this neighborhood, so we have

\begin{eqnarray}
 \text{degree}(v)\pi(v) = \sum_{u \sim v} \pi(u) \nonumber
\end{eqnarray}

from which we get the sum of $\pi(.)$ over any neighborhood in $V_{h}$ as,

\begin{eqnarray}
\pi(N(v)) = \pi(v) + \sum_{u \sim v} \pi(u) = (\text{degree}(v) + 1)\pi(v) \nonumber
\end{eqnarray}

The contribution of any local region in the restricted potential trace $\widehat{\Gamma}$ depends on just two possibilities regarding $h_{v}$. We deal with them separately.

\begin{itemize}
 \item[-] Case $h_{v} = 0$: $v \in I$, consequently none of its neighbors are in $I$. In this case the contribution to $\widehat{\Gamma}$ is the sum $\sum_{u \sim v}\pi(u) = \text{degree}(v)\pi(v) = \frac{\text{degree}(v)}{\text{degree}(v) + 1}\pi(N(v))$.

 \item[-] Case $h_{v} = 1$: $v \notin I$, and some of its neighbors are in $I$. Assume the worst case scenario when all the neighbors are in $I$. The contribution to $\widehat{\Gamma}$ in such a situation is just $\pi(v) = \frac{1}{\text{degree}(v) + 1}\pi(N(v))$.
\end{itemize}

It follows that any neighborhood $N(v)$ contributes at least $\frac{1}{\text{degree}(v) + 1}\pi(N(v))$ to $\widehat{\Gamma}$, regardless of the specific values of $h_{v}$. Since the degree is bounded by $\Delta$, this translates to a minimum contribution of $\frac{1}{\Delta + 1}\pi(N(v))$. Therefore, the value of $\widehat{\Gamma}$ in equation (\ref{equ:tclLower2}), for any independent set $I$, is at least

\begin{eqnarray} \label{equ:tclLower}
\frac{1}{(\Delta + 1)\pi_{w}(v)} \sum_{u}\pi_{w}(u)
\end{eqnarray}

The upper bound yielded by Lemma \ref{lem:tclUpp} is bounded from above by,

\begin{eqnarray} \label{equ:tclUpper}
\frac{(\Delta - 1)}{\pi_{w}(v)} \sum_{u}\pi_{w}(u)
\end{eqnarray}

Using equations (\ref{equ:tclUpper}) and (\ref{equ:tclLower}), we obtain the following two sided bounds for degree bounded graphs,

\begin{eqnarray} \label{equ:tclSandW}
\frac{1}{(\Delta + 1)\pi_{w}(v)} \sum_{u}\pi_{w}(u) \leq x_{v} \leq \frac{(\Delta - 1)}{\pi_{w}(v)} \sum_{u}\pi_{w}(u)
\end{eqnarray}

\begin{thm}\label{thm:consFac} For any sandpile with bounded vertex degrees, the minimum number of particles that need to be added at any vertex $v$ to observe a toppling at any vertex $w$ is equal, up to constant factors, to the following expression,
\begin{eqnarray}
\frac{1}{\pi_{w}(v)} \sum_{u}\pi_{w}(u)
\end{eqnarray}
\end{thm}

The computation of transience class $\text{tcl}(S)$ requires evaluating the above expression for all possible combinations of $v, w \in V(S)$. Computing the function $\pi(.)$ can be done in very efficiently following the recent path breaking work by \cite{KMP-1}, \cite{KMP-2}, \cite{ST} on solving symmetric, diagonally-dominant linear systems. Consequently, finding the pair with worst estimates is also easy to do.

\begin{cor} There exists a polynomial time algorithm which computes the transience class of a degree bound sandpile up to constant factors. \end{cor}

\section{The case of Grid Sandpile}
As noted in the introduction, the sandpile associated with the $n \times n$ grid is of particular importance. We define it formally below.

\begin{defin} Consider the $n \times n$ grid graph. Attach an extra sink node to the boundary such that there is a single edge to each non-corner boundary node and double edges to the corner nodes. We denote both the sandpile and the corresponding circuit by $\texttt{GRID}_{n}$.
\end{defin}

\noindent \textbf{Notation:} For the purposes of labeling the nodes, we assume the grid is embedded canonically in the first quadrant of $\mathbb{Z}^{2}$ with a corner coinciding with $(1,1)$. Every node on the grid is labeled with the coordinates it occupies in the lattice. The labels are $(i,j), 1 \leq i,j \leq n$. The sink node is labeled $s$.

Babai and Gorodezky \cite{LB07} have shown that $tcl(\texttt{GRID}_{n}) = O(n^{29.0095})$. In this section we will improve this bound to $O(n^{7})$. The following is a broad outline of our proof of Theorem \ref{thm:tclGrid}. We bound the potential profile $\Gamma(\texttt{GRID}_{n})$ and $\min_{v,w}\pi_{v}(w)$ separately and estimate the bound on $tcl(\texttt{GRID}_{n})$ using Theorem \ref{thm:tcl}. The bounds on $\Gamma(\texttt{GRID}_{n})$ are obtained using ideas based on charge conservation, along the lines of the classical Ampere's Law of electrodynamics. For bounding the value of $\min_{v,w}\pi_{v}(w)$, we show that values of $\pi_{v}(c)$ and $\pi_{w}(c)$ (where $c$ is the center) can be used to obtain estimates on $\pi_{v}(w)$. Using grid symmetries we prove monotonicity properties which imply that the minimum value of $\pi_{v}(c)$ is obtained when $v$ is a corner. Finally we bound the value of $\pi_{v}(c)$ by constructing a harmonic function with power applied at corner such that unit potential appears at the center. The construction of this distribution uses a certain potential domination property of the center over edges and the fact that the grid graph can be expressed as the Cartesian product of paths. The amenability of paths in constructing harmonic distributions and the classical superposition theorem (Theorem \ref{thm:sup}) play key roles in the construction. We begin in the next subsection, with the potential domination property.

\begin{subsection}{A potential domination property of the center}

% The circuit $\texttt{GRID}_{n}$ has very nice symmetry properties which enable one to infer non-trivial monotonicity properties when potential is applied at some special nodes. In this section,
We will consider the case when potential is applied at a corner and prove a kind of potential dominating property of the center over the corner opposite to the power source. The proofs of these monotonicity properties require concepts involved in proving convergence properties of iterative algorithms which solve boundary value problems. This procedure is known as the \textit{Jacobi Method} \footnote{See the wikipedia entry for the Jacobi method}. One starts with assigning the given values to boundary points and zero to every other node. In every iteration, the value of any internal node is updated according to the values of neighboring nodes just after the preceding iteration ended. When the linear system is irreducible weakly diagonally dominated (as in our case), it produces a set of values converging to the final solution. For a proof of convergence we refer the reader to \cite{GS}. This technique is folklore in basic finite element analysis and belongs to the much more general class of algorithmic constructions of solutions to Poisson's equation. The speed at which the values converge to the solution is intimately tied up to the rate of mixing on the underlying graphs. We will show that when potential is applied at a corner, the values that appear on the nodes in any iteration obey a simple monotonicity property, thereby implying that the solution (which is the point of convergence of these points) obeys the same monotonicity property.

\begin{defin}\textit{Corner Monotonicity}: Let $f$ be a function defined on a finite $n \times n$ grid, $f : \mathbb{Z}_{n}\times\mathbb{Z}_{n} \rightarrow \mathbb{N}$. We say $f$ is corner monotone with respect to $(1,1)$ if $f(p) \leq f(q)$ for any pair of lattice points $p$ and $q$ such that the segment $q-p$ is either perpendicular to the diagonal passing through $(1,1)$ or along one of the edges passing through it and $q$ is closer to the diagonal then $p$.
\end{defin}
Corner Monotonicity with respect to other corners is defined likewise , see figure (\ref{fig:cornerMon}).

\begin{figure}[ht]
\centering
\includegraphics[width=.6\textwidth]{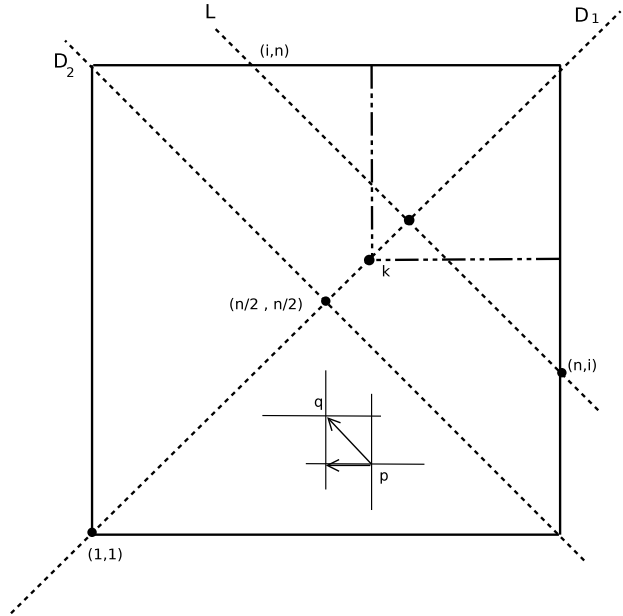}
\caption{Corner Monotonicity with respect to $(1,1)$}
\label{fig:cornerMon}
\end{figure}

Let $c_{0}$ be the starting set of values with $1$ assigned to $(1,1)$ and $0$ to every other node. Let $c_{t}$ be the set of values resulting from iteration number $t$. $c_{t+1}$ is obtained from $c_{t}$ using the following conditions of harmonicity of functions. For any node $v$,

\begin{eqnarray}\label{equ:toppInd}
 c_{t+1}(v) = \frac{\sum_{v' \sim v}c_{t}(v')}{deg(v)}
\end{eqnarray}

\begin{lem} If $c_{0}$ is corner monotone, then $c_{t}$ is corner monotone for all values of $t$.\end{lem}
\textbf{Proof:} Without loss of generality, assume that $q>p$ (the other possibility will be implied by symmetry). First consider the case when the segment $p-q$ is perpendicular to the diagonal through $(1,1)$. So if $q$ is of form $(x,y)$ then $p$ is $(x-1, y+1)$.

We will proceed by induction on the number of steps of the algorithm. Before the first iteration, time $t=0$, corner monotonicity of $c_{0}$ is trivial. Assume $c_{t}$ is corner monotone, we need to show that $c_{t+1}(q) \leq c_{t+1}(p)$. We now use equation (\ref{equ:toppInd}). Note that by induction hypothesis, each term in the expression of $c_{t+1}(q)$ is dominated by the respective term of $c_{t+1}(p)$ which implies $c_{t+1}(p) \leq c_{t+1}(q)$. A special case arises when $p$ lies on the diagonal itself. Here we make use of symmetry of the grid. When $p$ lies on the diagonal, its northern neighbor is mirror image of eastern neighbor and likewise for southern and western neighbors. The eastern and southern neighbors are common with $q$. The remaining two of $q$'s neighbors are dominated by these two. Again, by induction hypothesis, the inequality follows. The remaining reasoning is same as the standard case.

The other case of $p-q$ being parallel to an edge through $(1,1)$ edge is analogous. $\blacksquare$

The limiting value of $c$ is the harmonic distribution that results when a unit potential is applied at the node $(1,1)$. It satisfies the same monotonicity properties that the distributions $c(t)$ satisfied, for all values of $t$. This gives us the following lemma.

\begin{lem} When a potential is applied at a corner, then the resulting potential distribution is corner monotone with respect to that corner.\end{lem}

Using this, the following potential domination property of the center can be inferred.

\begin{lem}\label{lem:centOverEdge} When potential is applied at a corner, the potential at the center of the grid is higher than at any site on the opposite boundary.\end{lem}

\textbf{Proof:} Using corner monotonicity, we claim that when power is applied at node $(1,1)$ and unit potential is observed at some node $\{(n,i)\}$ on the opposite edge , then the center of the grid $(\frac{n}{2} ,\frac{n}{2})$, also has at least a unit potential. The reasoning behind this assertion is as follows. Because of symmetry, the site $(i,n)$ also has at least unit potential. On the line connecting these two sites, say $L$, the potentials first increase till one reaches the intersection with the diagonal $D_{1} : x=y$ and then decrease monotonically. This follows from the corner monotonicity lemma as the starting configuration is corner monotone.
So both $(1,1)$ and $L \cap D_{1}$(and in case  $L \cap D_{1}$ is not a lattice point, the two points closest to it) have at least unit potential. Assume there exists a point on the line segment joining $(1,1)$ to $L \cap D_{1}$, say $k$ whose potential is less then unity. Then every point to its right has potential less then unity, following corner monotonicity. Similarly for every point right above it. But this two sets partition the circuit into disjoint pieces, one of which contains $(1,1)$ and other contains $L \cap D_{1}$. Any random walk starting from $L \cap D_{1}$ and ending at $(1,1)$ has to pass through this set. The potential that appears on
$L \cap D_{1}$ cannot exceed the maximum value taken by any point in this set. This contradicts the assumption that potential at $L \cap D_{1}$ is greater then that at $k$. Hence such a $k$ cannot exist implying that every site on the line joining $(1,1)$ and $L \cap D_{1}$ has at least unit potential. $\blacksquare$

\noindent \textit{Note:} Lemma \ref{lem:centOverEdge} can be rephrased in the following manner. If applying a potential of $p(n)$ at a corner produces unit potential anywhere on an opposite boundary node, then applying $p(n)$ at any corner is enough to produce at least a unit potential at the center. We will later see an example of a harmonic distribution with a single positive pole at a corner and unit potential at some point on the opposite edge. The utility of this Lemma lies in the fact that in general constructing harmonic functions with an arbitrary pair of poles and known value at some arbitrary point is not easy. In our case, we need the potential that appears on a corner when potential is applied at the opposite corner. Our efforts so far, to construct a distribution with a pole at corner and known response at the opposite corner, have been fruitless. However, using Lemma \ref{lem:centOverEdge} in conjunction with the triangle inequality for potentials, we obtain fairly good estimates of the corner to corner potential correlations. We believe that the estimates we obtain are close to the square of the true value.

\end{subsection}

\begin{subsection}{The case of corner to corner}

We will now obtain a lower bound on the minimum value of $\pi_{v}(w)$ for any pair $v$ and $w$. We will show that minimum values of $\pi_{v}(w)$ are obtained when both $v$ and $w$ are points on the boundary of grid. Let the center of the grid be denoted by $c$. Then, using Lemma \ref{lem:potTri} (triangle inequality of potentials), we obtain $\pi_{w}(c)\pi_{c}(v) \leq \pi_{v}(w)$. Using Lemma \ref{lem:potRec},

\begin{eqnarray}
\pi_{w}(c) = \pi_{c}(w).\frac{R_{eff}(s,c)}{R_{eff}(s,w)} \nonumber
\end{eqnarray}

Clearly, $\beta \min_{v}\pi_{c}(v)^{2} \leq \min_{v,w}\pi_{v}(w)$, where $\beta$ is the minimum, upto constant factors, value of $\frac{R_{eff}(s,c)}{R_{eff}(s,w)}$ over all possibilities of boundary nodes $w$. The following lemma bounds the value of $\beta$.

\begin{lem}\label{lem:resRatio} The value of $\beta$ defined above is lower bounded by some constant. \end{lem} %$1/\log{n}$ .\end{lem}
\textbf{Proof}: The bound is derived in two parts. We first derive a lower bound on the numerator. Consider any node $w$. The effective resistance between sink node $s$, and $w$ decreases if we reduce any edge's resistance. This follows simply from the Reiligh's monotonicity principle. We reduce all the resistances, except the ones incident on $w$, to zero. This effectively leaves only node $w$ connected with $s$ by $4$ parallel edges. The net resistance of this configuration is $1/4$. This is an absolute lower bound on the effective resistance between \textit{any} node and sink (as the argument is independent of the location of node).

%Consider any boundary node $w$. The effective resistance between sink node $s$, which is adjacent to $w$, and $w$ decreases if we reduce any edge's resistance. This follows simply from the Reiligh's monotonicity principle. We reduce all the  resistances, except the ones incident on $w$, to zero. This effectively leaves only node $w$ connected with $s$ by $4$ parallel edges. The net resistance of this configuration is $1/4$. This is an absolute lower bound on the effective resistance between \textit{any} node and sink as the argument can be applied to any node.

An upper bound on the value of denominator follows from the fact that it is a parallel combination of a unit resistance with a network. The net resistance of parallel combination of $r_{1}$ and $_{2}$ is at most $\min \{r_{1}, r_{2}\}$.

%requires more work. We proceed by first deleting all the edges except the ones in the $n/2 \times n/2$ grid and the one connecting node $(1,1)$ to $s$. Again, the Reiligh's monotonicity principle ensures that this can only increase the value of $R_{eff}(c,s)$. We are left with a pair of nodes connected by unit resistance in series with a resistance of value $r_{g}$. Here $r_{g}$ is just the effective resistance between diagonally opposite corner nodes. From \cite{PB}, we know that this is $O(\log{n})$.

The above two facts give the required bounds on the value of $\beta$. $\blacksquare$

To show that $\pi_{c}(v)$ is minimum when $v$ is a corner node, we need another potential monotonicity lemma, the sandpile analogue of which appears in Babai and Gorodezky \cite{LB07}. However, the use of monotonicity properties in our proof is essentially different. Babai and Gorodezky \cite{LB07} use monotonicity along with the pigeonhole principle based combinatorial arguments to derive bounds on $tcl(\texttt{GRID}_{n})$. These arguments are first made on the infinite grid. Using monotonicity, \cite{LB07} bound the region which particles touch when they are added to single sites and by ensuring that the sizes of the regions are small, one can assume that the boundary is not touched and pretend to be on the infinite grid itself. Our use of monotonicity is much more straightforward in the sense that we want to find the pair of vertices with the worst estimates and monotonicity properties lead us directly to them.

\begin{defin}\textit{Center Monotonicity}: Let $f$ be a function defined on a finite $n \times n$ grid, $f : \mathbb{Z}_{n}\times\mathbb{Z}_{n} \rightarrow \mathbb{N}$. We say $f$ is center monotone if $f(p) \leq f(q)$ for any pair of lattice points $p$ and $q$ such that the segment $q-p$ is aligned perpendicular to some axis of symmetry and $q$ is closer to it then $p$.
\end{defin}

% \begin{lem} If $c_{0}$ is center monotone, then $c_{t}$ is center monotone for all values of $t$.\end{lem}

\begin{lem} When potential is applied at the center, then the resulting potential distribution is center monotone.\end{lem}

The proof of above lemma is completely analogous to the previous case. The center monotonicity lemma implies that \textit{if we apply a potential at the center, then the corner sites have the lowest potential} (among all non-sink nodes). Rephrasing in terms of reciprocals of $\pi(.)$, we get the following upper bound on the maximum value of $\pi_{v}(w)^{-1}$ over all pairs $w$ and $v$.

\begin{lem}\label{lem:chiHar} If applying the potential $p(n)$ on a corner induces unit potential at the center, applying $K.p(n)^{2}/\beta$ at any node induces unit potential at every non-sink node, where $K$ is a constant.
\end{lem}
% Since $\texttt{GRID}_{n}$ is regular (except for the sink node), using reciprocity lemma one can say that for inducing unit potential at center, the maximum potential that needs to be applied at any single site is for the corner nodes. Using this with lemma (\ref{lem:corToCor}) we get,

The only remaining information is the value, $p(n)$, of the potential that when applied at a corner, induces a unit potential at the center. In the next section we will see an example of a such a harmonic distribution.

\noindent \textit{Note:} In the preceding discussion, we have conveniently assumed that $n$ is odd, else no such center site would exist. It is however easy to extend the discussion to the case of even $n$.
\end{subsection}

\begin{subsection}{Constructing a harmonic distribution over $\texttt{GRID}_{n}$: determining the corner to center response}

% The fundamental question arising in potential theory is of the following type; when unit potential is applied at some node, what is the potential appearing at some other node? Which means evaluation or estimation of the function $_{s}\pi_{t}(.)$. This is the central theme of discussions related to the Harnack inequality over graphs. We are concerned with the grid based circuit described above and wish to calculate the potential one needs to apply at $(n,n)$ so that unit potential appears at $(1,1)$. Evaluating this number directly is somewhat tedious as the harmonic distribution when potential is applied at the corner is not describable using simple polynomial functions. The exact form of this distribution will be discussed in detail in following sections.

Our current goal is to construct a harmonic distribution with power applied at a corner such that at least unit potential appears at the center, or in other words bound the value $\pi_{v_{corner}}(c)$. In general, constructing harmonic distributions with arbitrary poles and known values at some node in general is difficult. However, in our present problem, we will bound this quantity using the fact that the grid is the \textit{Cartesian product} of paths and that potential functions on paths are easy to construct.

Consider the (path) line circuit which has $n$ nodes. The last node is connected to ground potential through a unit resistance. We apply a potential of $n+1$ units \textit{through} a unit resistor at node labeled $n$ and observe that unit potential appears at the corner vertex (labeled $1$). See figure (\ref{fig:lineCirc-2}).

\begin{figure}[ht]
\centering
\includegraphics[width=.7\textwidth]{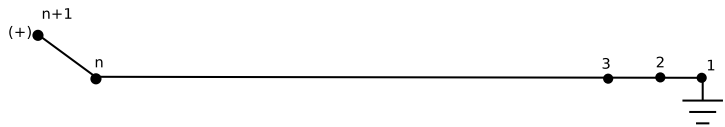}
\caption{Line Circuit}
\label{fig:lineCirc-2}
\end{figure}

\noindent \textit{A harmonic distribution on $\texttt{GRID}_{n}$:} Now take the $n \times n$ grid with ground connection attached to each of its boundary nodes on the left and bottom edges through unit resistances. Power sources are applied at the top and right edges through unit resistances. We apply a potential of $(n+1).i$ at the boundary nodes $(n,i)$ and $(i,n)$. At the special corner node $(1,1)$, we apply $n^{2} + n$. One can check that the potential that appears at any grid node $(i,j)$ is $V(i,j) = i.j$. In particular, unit potential appears at node $(1,1)$, i.e. $V(1,1) = 1$. This construction is a particular case of constructing a harmonic distribution on the Cartesian product of two graphs given a harmonic distribution on each of them. The generalization is discussed in the full version.

Using the superposition principle (Theorem \ref{thm:sup}), the potential value at $(1,1)$ due to these $2n-1$ power sources is the sum of potential values that would have appeared when these power sources would have been used one at a time with all other sources short circuited, at their respective positions. Also, among all the nodes on the top and right edges, there exists one with the maximum potential response at $(1,1)$, i.e. where when unit potential is applied, the potential at $(1,1)$ is maximum. Again using superposition principle, if all the power sources are applied at this site alone with all other sites connected to sink, at least unit potential appears at $(1,1)$. The value of this new power source is $n^{2} + n + \sum_{i=1}^{n-1}2.(n+1).i = n^{3} + n^{2} = O(n^{3})$. However, the site on which power source is applied has exactly one connection less with sink compared to the circuit $\texttt{GRID}_{n}$ in which we apply power through unit resistances. To remedy this, we add an extra edge to the sink. One can check that this modification is non-essential and one needs to change the power applied at this corner by the amount of current flowing this new edge, which amounts to an at most constant factor change in the power applied. The same potentials appear at other nodes. Using reciprocity and (degree) regularity of ordinary sites in $\texttt{GRID}_{n}$ one can interchange this power node on an edge and the corner node $(1,1)$ to obtain the following lemma.

% Using symmetry and superposition principle, one can apply twice the previous power sources at the node $(i,n)$ instead of one each at $(i,n)$ and $(n,i)$ without changing the value of $V(1,1)$, for all values of $i$. We now have power sources only on the top edge and all the nodes on the right edge are now connected to the sink node. The potential value $V(1,1)$ due to these $n$ power sources is the sum of potential values that would have appeared when these power sources would have been used one at a time, at their respective positions. Also, among all the sites on the top edge, there exists a site, where when unit potential is applied, the potential $V(1,1)$ is maximum. If all $n$ power sources are moved to that site, then $V(1,1) \geq 1$. The value of this new potential is therefore $n^{2} + n + \sum_{i=1}^{n-1}2.(n+1).i = n^{3} + n^{2}$. Therefore we have the following:
%
% \begin{lem} In $\texttt{GRID}_{n}$, there exists a point on the top edge, where applying $O(n^{3})$ potential induces at least unit potential at $(1,1)$\end{lem}
%
% Using the potential reciprocity lemma (Lemma \ref{lem:potRec}), one can turn this around to obtain,

\begin{lem}\label{lem:corToEdge} In $\texttt{GRID}_{n}$, applying $O(n^{3})$ potential at $(1,1)$ induces $O(1)$ potential at some point on the top edge. \end{lem}

Using Lemmas \ref{lem:corToEdge} and \ref{lem:centOverEdge}, we get the following result:

\begin{lem}\label{lem:corToCen} In $\texttt{GRID}_{n}$, applying $O(n^{3})$ potential at $(1,1)$ induces at least $O(1)$ potential at the center. \end{lem}

\noindent \textit{Remark (Improving the lower bounds on $tcl(\texttt{GRID}_{n})$)}: Lemma \ref{lem:corToEdge} observes a pair of a vertices, both at boundary, such that applying $O(n^{3})$ potential at one of them induces at least unit potential at the other. The paragraph preceding this Lemma outlines the proof of this property by shifting the power sources to the \textit{best response} point on the edge. If however one shifts these sources to the \textit{worst response} point, the existence of a complementary vertex, with respect to the corner, can be proved. This pair has the property that applying $O(n^{3})$ potential at one of them induces at most unit potential at the other. Using the Lemma \ref{lem:tclsil}, we obtain that the number of particles that can be added at one of them without toppling the second one is lower bounded by $\Omega(n^{3})$. This is an improvement over the obvious lower bounds of $\Omega(n^{2})$.

\begin{cor}$tcl(\texttt{GRID}_{n}) = \Omega(n^{3})$.\end{cor}

\end{subsection}

\begin{subsection}{Bounding the potential profile of $\texttt{GRID}_{n}$}

Using the discussion preceding Lemma \ref{lem:tcl} we can get a bound of $O(n^{2})$ for $\Gamma(\texttt{GRID}_{n})$, which yields an $O(n^{9}\log{n})$ bound on $tcl(\texttt{GRID}_{n})$. Here we improve the bound on $\Gamma(\texttt{GRID}_{n})$ to $O(n)$ using current conservation arguments and the regularity of normal nodes of $\texttt{GRID}_{n}$.
Using property (\ref{prop:boundSand1}), one knows that the last site to topple is always at the boundary. Hence, when using theorem \ref{thm:tcl} for estimating transience classes, we know that the current source will always be added to a boundary node (adjacent to the sink). Consider our network, $\texttt{GRID}_{n}$, with a current source attached to some node $v$ adjacent to the sink, such that the potential of $v$ is unit.
 %As in the previous discussion, the sink node is maintained at zero potential. Consequently the potential at every other node is positive.
Note that the total current flowing in, $i$, is bounded above by the degree of $v$ ($4$ in this case). To see this, consider the equation of current conservation at $v$. %The current inflow is equal to current outflow.
$v$ is at unit potential and each of the neighbors' potential is non-negative. Consequently, at most a unit of current flows through each incident edge. So the total outflow is bounded from above by $4$ units. Hence, the total inflow from the current source is also bounded from above by $4$ units. Thus, applying a current source of $O(1)$ units produces a unit potential at the node $v$. Using this fact, we can prove the following bound on $\Gamma_{v}(\texttt{GRID}_{n})$.

\begin{lem}\label{lem:potProGrid} Consider any vertex $v$ on the boundary of grid in $\texttt{GRID}_{n}$. The potential profile $\Gamma_{v}(\texttt{GRID}_{n})$ induced due to $O(1)$ current source at $v$ is $O(n)$.\end{lem}

\textbf{Proof}: Denote the potential at node $v$ by $\pi(v)$. The total current going into the sink node is, say, $i$ and is equal to
\begin{eqnarray}
 i & = & \sum_{(v,s) \in E(\texttt{GRID}_{n})}\pi(v) \nonumber
\end{eqnarray}

The current going in to the sink is a sum above all the constituting currents through each of the incident edges. Since the potential of $s$ is zero, each of these currents is equal in magnitude to the potential of the neighboring nodes. If the set of nodes on corners are denoted by $C_{n}$, those on the interior of the boundary edges by $I_{n}$, and the union of these two by $B_{n}$, we can rewrite the value of $i$ in the following form.

\begin{eqnarray}\label{equ:netCurrIn}
 i &=& 2.\sum_{v \in C_{n}} \pi(v) + \sum_{v \in I_{n}} \pi(v)
\end{eqnarray}

Since all potentials are positive, the following inequality follows from equation (\ref{equ:netCurrIn}).

\begin{eqnarray}\label{equ:netCurr}
 i &\geq& \sum_{v \in B_{n}} \pi(v)
\end{eqnarray}

Now consider the sequence of smaller $(n-2k) \times (n-2k)$ concentric grids nested in the larger $n \times n$ grid. Define the sets $C_{n-2k}$, $I_{n-2k}$ and $B_{n-2k}$ analogously for each of these. For any element $v \in I_{n-2(k+1)}$, denote by $n(v)$ the unique neighbor lying in $I_{n-2k}$, and for a $v \in C_{n-2(k+1)}$, denote  the two neighbors by $n_{1}(v)$ and $n_{2}(v)$. Then, for each of these smaller grids, the net current entering through the set $B_{n-2k}$ is zero. In terms of potential functions, the condition can be stated as

\begin{eqnarray}
 \sum_{v \in C_{n- 2(k+1)}} (2.\pi(v)-\pi(n_{1}(v)) - \pi(n_{2}(v))) + \sum_{v \in I_{n - 2(k+1)}}( \pi(v) - \pi(n(v))) &=& 0 \nonumber
\end{eqnarray}

Separating the vertices belonging to boundaries of different grids, we obtain

\begin{eqnarray}
 2\sum_{v \in C_{n- 2(k+1)}} \pi(v) + \sum_{v \in I_{n - 2(k+1)}}\pi(v) &=& \sum_{v \in I_{n - 2k}}\pi(v)
\end{eqnarray}

Again using the fact that all potentials are positive, we get

\begin{eqnarray}\label{equ:gridBounds}
 \sum_{v \in B_{n- 2(k+1)}} \pi(v) &\leq& \sum_{v \in B_{n - 2k}}\pi(v)
\end{eqnarray}

Every vertex belongs to the boundary of exactly one concentric grid. Using equation (\ref{equ:netCurr}) and (\ref{equ:gridBounds}), we get

\begin{eqnarray}
 \sum_{v \in \texttt{GRID}_{n}} \pi(v) &\leq& \frac{n}{2}\sum_{v \in B_{n}}\pi(v) = \frac{ni}{2}
\end{eqnarray}

Since the degree of normal vertices is $4$, we have the following bound on the potential profile of grid when power is applied at some vertex of boundary $B_{n}$.

\begin{eqnarray}
 \Gamma_{B_{n}}(\texttt{GRID}_{n}) = 4\sum_{v \in \texttt{GRID}_{n}} \pi(v)  &\leq&  2ni
\end{eqnarray}
 $\blacksquare$

\end{subsection}

\begin{subsection}{Transience Class of $\texttt{GRID}_{n}$: a new bound}\label{sec:gridBound}

Using Lemmas \ref{lem:chiHar}, \ref{lem:corToCen} and \ref{lem:resRatio}, we obtain the following result which bounds the value of $\max_{v,w}(\pi_{v}(w))^{-1}$ from above.
%The following lemma, which bounds the rate of growth of any harmonic function over $\texttt{GRID}_{n}$ with sink as one pole and any other node as the second pole, is obtained.

\begin{lem}\label{lem:gridHarn}In $\texttt{GRID}_{n}$, applying $O(n^{7})$ potential at any site induces at least $O(1)$ potential everywhere. \end{lem}
%\begin{lem}\label{lem:gridHarn}In $\texttt{GRID}_{n}$, applying $O(n^{7}\log{n})$ potential at any site induces at least $O(1)$ potential everywhere. \end{lem}

Using Lemma \ref{lem:gridHarn}, \ref{lem:potProGrid} and Theorem \ref{thm:tcl} we have the following bounds on $tcl(\texttt{GRID}_{n})$.

\begin{thm}\label{thm:tclGrid}$tcl(\texttt{GRID}_{n}) = O(n^{7})$.\end{thm}
%\begin{thm}\label{thm:tclGrid}$tcl(\texttt{GRID}_{n}) = O(n^{7}\log{n})$.\end{thm}

\noindent \textit{Remark:} While the bounds proved above mark a substantial improvement over the current known $O(n^{29.0095})$, experiments suggest a bound of somewhere $O(n^{4})$. The estimates on the value of potential profile has little scope of improving substantially. Constructing the harmonic distribution with more care seems to be a plausible approach. Another possibility lies in exploiting the planarity of the sandpile graph. We will explore this avenue in further detail in the next section and obtain closed form expressions on bounds of $tcl(S)$, when $S$ is planar, in terms of the spectrum of the Laplacian of the dual of $S$.

\end{subsection}

\section{The case of planar Sandpile}

After showing the intimate relationship between the transience class of a sandpile and the harmonic functions over the underlying graphs, we will now show that if the underlying graphs are planar, the bounds on transience class can be expressed in a much more explicit algebraic form. Consider the sandpile $S$ and the corresponding circuit, both of which will be assumed to be planar.

In the circuit $S$, we apply a unit potential \textit{across} some boundary edge and observe the potential at some boundary node (as has already been noted, boundary nodes suffice for our worst case analysis). Now take the dual planar circuit of $S$, say $\widetilde{S}$ (for a detailed discussion of dualising operations in context of harmonic functions, see Benjamini and Lovasz \cite{BL03}). For every edge in the original graph, there exists exactly one edge in the dual graph. Call these edges dual of each other. There is a special edge in the circuit, the \textit{power edge}, across which the potential source is attached. Its dual edge becomes the unit current source in the dual circuit. The potentials at nodes in original circuit satisfied the Kirchoff's current law (the condition of harmonicity of voltages). The potential difference across each edge becomes the current flowing through the respective dual edge in the dual graph. And equations of Kirchoff's current law become those of loop law in the dual. Since all the currents satisfy the loop law, the potentials thus developed satisfy the current law as well. The estimation of potential difference across any edge in the first circuit is equivalent to estimating the current across the dual edge in the dual circuit. For any boundary vertex, its potential difference with that of sink equals in magnitude the current through the boundary edge incident on this node. The same current flows through the dual edge in the dual circuit. So, estimating the potential of a boundary node in the original graph is equivalent to estimating some current in the dual graph.

\begin{figure}[ht]
\centering
\includegraphics[width=.7\textwidth]{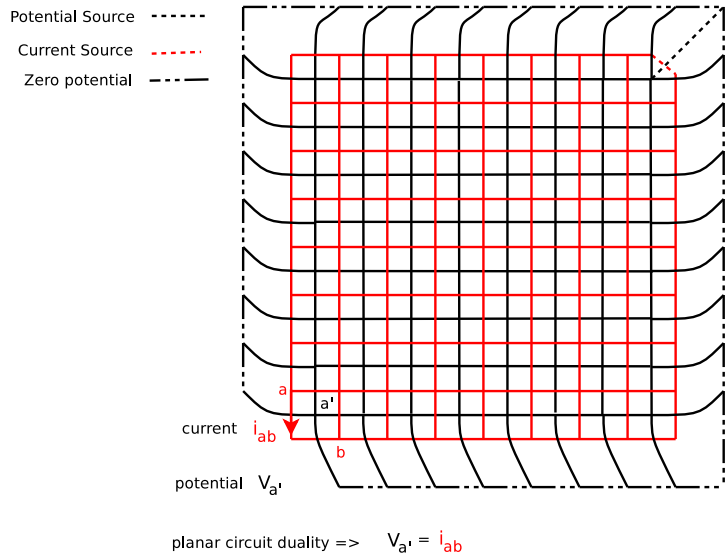}
\caption{Grid Circuit with its dual}
\label{fig:gridDual}
\end{figure}

As an example of interest, consider the grid graph with sink attached to border. We take its dual graph. See figure (\ref{fig:gridDual}). The original circuit is shown in black lines and the dual in red lines. The dotted edges on the top right corner are the power sources of the two circuits. If for a unit potential applied at top right corner produces $x$ at the bottom left corner (labeled $a'$), then unit current source through the top right edge produces a current $x$ through the bottom left resistor (labeled $ab$).

\noindent \textbf{Note}: The current source is placed \textit{across the dual edge}. Which means that if the dual edge is connecting the nodes $u$ and $v$, such that source is attached to $u$ and sink to $v$, then a unit current flows \textit{from $v$ to $u$} in the edge $uv$, to maintain flow conservation equations. In network theory terms, the edge $uv$ itself is the current source, and as such its internal flows must not be taken into account while writing the Kirchoff's equations. Consequently we delete the edge $uv$ and simply attach a current source at $u$ and sink at node $v$. The graph obtained after deleting edge $uv$, dual to edge $e$ in $S$, is called the \textit{restricted dual} of $S$ and denoted by $\widetilde{S}_{e}$. We will use the same symbol to denote the underlying graph. The current source is attached to vertex $u$ and sink to vertex $v$. Let the combinatorial Laplacian of this graph be denoted by $L$, the potentials that appear at each of the nodes because of the current flowing by vector $Z$ and let $I$ be the vector containing \textit{net currents} flowing in at any node. The equations of Kirchoff's laws at each node can be succinctly written as,

\begin{eqnarray}
LZ = I \nonumber
\end{eqnarray}

The vector $I$ has all entries $0$ except for a $1$ at position corresponding to node $u$ and $-1$ for node $v$. Given the values of $L$ and $I$, we need to estimate the potential difference between nodes $p$ and $q$, or equivalently, the current in the edge $pq$. Because matrix $L$ is singular, it is not possible to resolve the question by the usual methods of estimating certain entries of the inverse matrix. However, using the fact that $L$ is symmetric one can indeed \textit{almost} invert it enough to suffice for our purpose. See, for instance the paper by Wu \cite{Wu04} to compute two point resistances in networks.

\begin{lem}\label{lem:lapCurr}Consider a resistive network $S$ with Laplacian $L$ whose eigenvalues are  $\lambda_{0} = 0 < \lambda_{1} \leq \ldots \lambda_{n-2} \leq \lambda_{n-1}$ and  $\Psi$ is the unitary matrix containing the eigenvectors. The $i^{th}$ column, $\psi_{i}$, is the eigenvector corresponding to $\lambda_{i}$. If unit current is injected at node $u$ and taken out from node $v$ then the magnitude of current in edge $pq$ is given by
 \begin{eqnarray}\label{equ:currSize}
 i_{pq} \qquad = \qquad \vert \sum_{\substack{0 < k \leq n-1 }} \frac{(\psi_{k}(p) - \psi_{k}(q))(\psi_{k}(u) - \psi_{k}(v))^{\dag}}{\lambda_{k}} \mid
\end{eqnarray}
\end{lem}

\textbf{Proof }: Denote by $L(\epsilon)$ the matrix $L + \epsilon I$. Note that for $\epsilon > 0$, $L(\epsilon)$ is invertible, unlike $L$. Call its inverse $G(\epsilon)$. Denote the row of $G(\epsilon)$ corresponding to node $a$  by $G(\epsilon)(a)$.

\begin{eqnarray}
 Z(a) & = & \lim_{\epsilon \rightarrow 0}( G(\epsilon)(a).I ) \nonumber
\end{eqnarray}

Knowing the value $I$ explicitly, we can write

\begin{eqnarray}\label{equ:potDiff}
Z(a) = \lim_{\epsilon \rightarrow 0}(G(\epsilon)(a,u) - G(\epsilon)(a,v))
\end{eqnarray}

where $G(\epsilon)(x,y)$ is the entry in the row of the node $x$ and column of node $y$. The formula for $G(\epsilon)$ is,

\begin{eqnarray}
G(\epsilon) = \Psi \Lambda(\epsilon)^{-1} \Psi^{\dag} \nonumber
\end{eqnarray}

where $\Psi$ is the unitary matrix containing the eigenvectors of $L$ (and consequently of $L(\epsilon)$ and $G(\epsilon)$) as its columns, $\Psi^{\dag}$ is its hermitian and $\Lambda(\epsilon)$ is the diagonal matrix containing the eigenvalues of $L(\epsilon)$. Exactly one of these eigenvalues is $\epsilon$ (for connected graphs). The corresponding eigenvector has every entry $1/n$. Let the eigenvalues be $\lambda_{0} = \epsilon < \lambda_{1} \leq \ldots \lambda_{n-2} \leq \lambda_{n-1}$. We obtain the following expression for $G(\epsilon)(x,y)$.

\begin{eqnarray}\label{equ:pot}
G(\epsilon)(x,y) = \frac{1}{n^{2}\epsilon} + \sum_{\substack{0 < k \leq n-1 }} \frac{\psi_{k}(x)\psi_{k}(y)^{\dag}}{\lambda_{k} + \epsilon}
\end{eqnarray}

where $\psi_{k}(x)$ is the entry of node $x$ in the $k^{th}$ eigenvector. Using (\ref{equ:potDiff}) and (\ref{equ:pot}) we get,

\begin{eqnarray}
 Z(a) = \sum_{\substack{0 < k \leq n-1 }} \frac{\psi_{k}(a)(\psi_{k}(u) - \psi_{k}(v))^{\dag}}{\lambda_{k}}
\end{eqnarray}

The amplitude of the current in the edge $e$ between nodes $p$ and $q$ is equal in magnitude to $Z(p) - Z(q)$. That is,

\begin{eqnarray}
 i_{pq} \qquad = \qquad \vert \sum_{\substack{0 < k \leq n-1 }} \frac{(\psi_{k}(p) - \psi_{k}(q))(\psi_{k}(u) - \psi_{k}(v))^{\dag}}{\lambda_{k}} \mid \nonumber
\end{eqnarray}
$\blacksquare$

In the circuit $S$, the possible locations of a power source are just the set nodes connected to the sink. The possible current source and sink nodes in the dual $\widetilde{S}$ are exactly the ones corresponding to dual (boundary) edges in $S$. Let $\{ e_{i} \}_{i = 0}^{k}$ be the set of boundary edges sandpile $S$. Denote by $_{k}i_{pq}$, the current in edge $pq$ in $\widetilde{S}_{e_{k}}$ when current source is applied across the edge dual to $e_{k}$ in $S$. Using this notation with Lemmas \ref{lem:tcl} and \ref{lem:lapCurr}, we can bound the transience class of $S$ purely in terms of eigenvalues and eigenvectors of the Laplacians of restricted duals of $S$.

\begin{lem} For planar sandpile $S$, $tcl(S) = |S|.O(\max_{k}\max_{pq}|_{k}i_{pq}^{-1}|)$.\end{lem}

\textbf{Proof}: The quantity $\min_{pq}|_{k}i_{pq}|$ gives the smallest boundary node potential value when the power edge in the original graph $S$ is $e_{k}$. Further, $\min_{k}\min_{pq}|_{k}i_{pq}|$ gives the minimum boundary node potential value over \textit{all} possible placements of power source. The reciprocal of this value is equal to the quantity $\max_{v,w}\{\pi_{w}(v)^{-1}\}$. The potential profile of any circuit is bounded by the size of the graph. $\blacksquare$

% OLD VERSION OF PREVIOUS PARAGRAPH
% In the circuit $S$, the possible locations of a power source are just the set nodes connected to the sink. The possible current source and sink nodes in the dual $\widetilde{S}$ are exactly the ones corresponding to dual (boundary) edges in $S$. Let $\{ e_{i} \}_{i = 0}^{k}$ be the set of edges dual to the boundary edges of the sandpile $S$. Denote by $_{k}i_{pq}$, the current in edge $pq$ in $\widetilde{S}$ when in the dual $S$, power is applied across edge $e_{k}$. $\min_{pq}|_{k}i_{pq}|$ gives the smallest boundary node potential value when the power edge in the original graph $S$ is $e_{k}$. Further, $\min_{k}\min_{pq}|_{k}i_{pq}|$ gives the minimum boundary node potential value over \textit{all} possible placements of power source. The reciprocal of this value is equal to the quantity $\max_{v,w}\{\pi_{w}(v)^{-1}\}$. Using this with Lemma \ref{lem:tcl}, we obtain bounds on the transience class of $S$ purely in terms of eigenvalues and eigenvectors of the Laplacians of restricted duals of $S$.

\noindent \textit{Remark:} Note that one could as easily invert the toppling matrix of the sandpile to work out a similar formula. Being the principal minor of a connected graph's Laplacian, it is invertible. But as opposed to Laplacians, their principal minors are not well explored. While we know the eigenfunctions of Laplacians of most common classes of graphs, the same cannot be said about the principal minors. It is in this light that the above simplification becomes important. Its power can be displayed by considering the example of the grid. We discuss this example further in the last section on future work.

\section{Sandpile with $k$ connections to the sink}
We will illustrate the power of our reduction by proving the following theorem for any sandpile $S$ having at most $k$ connections to sink.

\begin{thm}\label{thm:ksink}
 $tcl(S)$ is $O((|S|+2)^{k}) $
\end{thm}

In a sandpile $S$, let the vertices connected to the sink be labeled $v_{1}, v_{2}, \ldots, v_{k}$. We have already seen that the worst case estimates of $\max_{v,w}\{\pi_{w}(v)^{-1}\}$ are obtained, when both the nodes labeled $v$ and $w$ are on the boundary, i.e. when both are directly connected to the sink. Henceforth, we will limit our discussion to only those pairs of sites which are from this boundary set $B_{S} = \{v_{i}\}_{i=1}^{k}$. So, $tcl(S)$ is bound from above by $O(|E(S)|. \max_{1 \leq i,j \leq k}\{\pi_{v_{i}}(v_{j})^{-1}\})$.

%\begin{eqnarray} \label{equ:tclksink1} %\end{eqnarray}

\begin{figure}[ht]
\centering
\includegraphics[width=.7\textwidth]{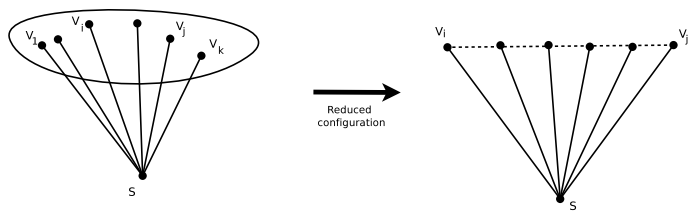}
\caption{Replacing the circuit with a reduced circuit}
\label{fig:circEqui}
\end{figure}

Consider the circuit $\widehat{S}$. The sink site, $s$, has $k$ edges incident on the remaining graph as shown in figure(\ref{fig:circEqui}). Assume that among all ordered pairs of vertices $(v,w)$ from $B_{S}$, the value of $\pi_{w}(v)$ is minimum for the pair $(v_{i},v_{j})$. We will obtain lower bounds on the value of $\pi_{v_{i}}(v_{j})$ in terms of size of $S$ and $k$, the number of connections with the sink.

We have already mentioned that $\pi_{v_{i}}(v_{j})$ is the potential that appears at node $v_{j}$ when unit potential is applied at node $v_{i}$. This is equivalent to saying that the probability of a random walk starting at node $v_{j}$ hitting node $v_{i}$ before the sink is equal to $\pi_{v_{i}}(v_{j})$. Consider the graph $S - \{s\}$ obtained from $S$ by deleting $s$ (this is the induced sub-graph over the set $V_{o}$). In this graph, if there is no path from node $v_{i}$ to $v_{j}$ then $\pi_{v_{i}}(v_{j})$ is zero, consequently its transience class is infinite. This refers to the condition that the sandpile graph must stay connected even if we remove the sink, for the transience class problem to make sense, as noted by Babai and Gorodezky \cite{LB07}. If every path between some pair of nodes passes through the sink, then one can add an unbounded number of particles at one of these sites without being able to induce a toppling at the other site. The following lemma summarizes this observation.

% \begin{lem}Deleting a bridge edge of $S - \{s\}$ creates a pair of sites $v,w$ for which $\pi_{v}(w)$ is zero.\end{lem}

Our goal is to construct a sequence of graphs $\{S_{p}\}_{0}^{m}$ with $S_{0} \equiv S$ and having some nice monotonicity properties on the value of $\pi_{v_{i}}(v_{j})$. In $S_{0} -\{s\}$, consider any edge, $e$, whose deletion does not disconnect $v_{i}$ and $v_{j}$. We apply a unit potential at site $v_{i}$ and observe the potential at node $v_{j}$. For the edge $e$, either reducing its resistance (eventual contraction) or increasing its resistance (eventual deletion) decreases $\pi_{v_{i}}(v_{j})$. This fact follows from a trivial extension of exercise $II.4.15$ in Bollob\'{a}s \cite{Bol}. $S_{1}$ is obtained from $S_{0}$ by deleting/contracting $e$, whichever operation reduces $\pi_{v_{i}}(v_{j})$. For the sake of uniformity, the node labels of $S_{1}$ are inherited from $S_{0}$ canonically. When an edge is contracted, the resulting node can be labeled with the label of any of the colluding nodes. Note that at no time nodes $v_{i}$ and $v_{j}$ get merged as this would increase $\pi_{v_{i}}(v_{j})$ to $1$, which contradicts the property that in any iteration, its value can not increase.

Similarly, in the $p^{th}$ iteration, $S_{p-1}$ is transformed to $S_{p}$ by picking any edge of $S_{p-1}-\{s\}$ whose deletion does not disconnect $v_{i}$,$v_{j}$ and contracting/deleting as appropriate. Observe that in every iteration, the total number of edges goes down by one. So the algorithm terminates with some graph $S_{m}$ which satisfies the following property.

\begin{prop} If $S_{m}$ is the graph obtained after the last iteration, deleting any edge in $S_{m} - \{s\}$ disconnects $v_{i}$ and $v_{j}$.\end{prop}

Which is equivalent to saying that the graph $S_{m} - \{s\}$ is just the path from $v_{i}$ to $v_{j}$, depicted by the circuit on right in figure (\ref{fig:circEqui}). Note that none of the edges incident to the sink get deleted/contracted during the whole process. Naturally, the path length is bound from above by the size of $E(S)-k$.

% For each of these non-bridge edges, either contracting them makes the potential observed at node $j$ go down or deleting them. Consider these edges in any order and repeatedly perform the following operation. If deleting the edge under consideration reduces $\pi_{v_{i}}(v_{j})$ then delete it, otherwise contract it. Note that in every iteration,  while the set of bridge edges grows. At the start of every iteration, we consider the edges that are non-bridge in the graph obtained after the last iteration.

% Henceforth, we will refer to the vertices on the path $v_{i} \rightarrow v_{j}$ as

Now consider the circuit $S_{m}$. We will be observe only the adjacent nodes of $s$. The resistance connecting any two of these (which occur consecutively on the path $S_{m} - \{s\}$) is bounded above by the total path length, $|E(S)| - k$. Unit potential is applied at node $v_{i}$ and the node $s$ is maintained at zero potential. The embedding shown in figure(\ref{fig:circEqui}) demonstrates that the circuit is planar. Before going further, we will need some basic properties of planar circuits in which power is applied across a boundary edge (note: \textit{boundary} is used here in the usual context of planar graph embeddings and should not be taken to mean adjacency with the sink node).

\begin{figure}[ht]
\centering
\includegraphics[width=.4\textwidth]{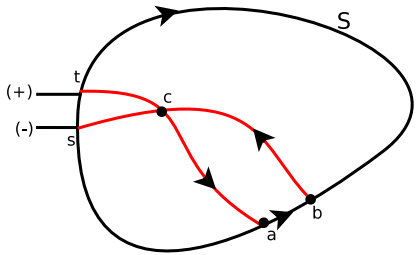}
\caption{Direction of current in the boundary}
\label{fig:boundCurrDir}
\end{figure}

% \begin{lem}\label{lem:boundCurrDir} Let $S$ be a planar resistive circuit with a given embedding such that the power source is attached across an edge on the boundary, say to vertices $t$ (positive) and $s$(negative). There are exactly two paths from $t$ to $s$ along the boundary. Then the direction of current in any boundary edge is along these paths from $t$ to $s$.\end{lem}

\begin{lem}\label{lem:boundCurrDir} Let $S$ be a planar resistive circuit with a given embedding such that the power source is attached across a pair of vertices on the boundary, say $t$ (positive) and $s$(negative). There are exactly two paths from $t$ to $s$ along the boundary. Then the direction of current in any boundary edge is along these paths from $t$ to $s$.\end{lem}

\textbf{Proof} : Assume there exists an edge $(a,b)$, in which the current flows from $a$ to $b$, i.e. the direction opposite to the path from $s$ to $t$. Since net inflow of current occurs only at node $s$, starting from $a$ one can construct a path to node $s$ such that in each edge current flows from $s$ to $a$. Similarly one can construct a path from $b$ to $t$ where in every edge current flows from $b$ to $t$. Because of planarity, these paths should intersect at some node, say $c$. Then we have a loop (as shown in figure (\ref{fig:boundCurrDir})) from $a$ to $b$ to $c$ to $a$ in which current flows in the same orientation in every edge. This contradicts Kirchoff's theorem about sums of potential differences along any loop in a circuit. Hence such an edge $(a,b)$ cannot exist. $\blacksquare$

This gives us the direction of current in every boundary edge, given a planar circuit (with some embedding) and a power source applied across a boundary edge. We now consider the effect of changing the resistance of some boundary edge on the potentials that appear on the nodes along the boundary. The following lemma tells us the change required in the resistance of a boundary edge to induce the desired effect on the potentials elsewhere on boundary.

% OLD STATEMENT
% In particular, if we wish to reduce the potentials that appear on the boundary nodes, should we reduce or increase the resistance of a particular edge on boundary. The second lemma answers this question and allows us to change resistances of boundary edges with the desired effect on the potentials elsewhere.

\begin{lem}\label{lem:boundResVar} Let $S$ be a planar resistive circuit with a given embedding such that the unit potential is applied across an edge on the boundary, say node $s$ (positive) and the sink $t$ is maintained at zero potential. Given any boundary edge $e$, increasing its resistance will decrease the potential that appears on any node along the portion of boundary between $e$ and $t$ and increase the potentials for the portion lying between $s$ and $e$.\end{lem}

\textbf{Proof} : Using lemma \ref{lem:boundCurrDir}, we know that in any edge, say $e$, along the boundary, current flows in the direction of the boundary path from $s$ to $t$. If we \textit{increase} the resistance of $e$, i.e. $dZ_{e} > 0$ (where $Z_{e}$ is the resistance of $e$), the effect on current flowing through any other boundary edge can be predicted using the compensation theorem. Previously current $I_{e}$ was flowing in direction $s$ to $t$ and $dZ_{e}$ is positive. The power source of $-I_{e}dZ_{e}$ when inserted in $e$, induces a current in the direction $t$ to $s$ (again using lemma \ref{lem:boundCurrDir}). Hence, the effect of increasing the resistance of $e$ is that current in every boundary edge decrease. Since potential of any boundary node,say $v$, between $e$ and $t$ is simply $\sum Z_{e'}I_{e'}$, where the sum is over all edges lying between $t$ and $e$. Since the resistances are constant and currents are decreasing, the sum also goes down. The case of vertex $v$ lying between $s$ and $e$ is analogous. Except for the fact that the potential of $v$ in this case is $1 - \sum Z_{e'}I_{e'}$. Increasing the resistance $Z_{e}$ decreases the summation (like in previous case) and so the net value increases. This completes the proof of lemma. $\blacksquare$

\begin{figure}[ht]
\centering
\includegraphics[width=.65\textwidth]{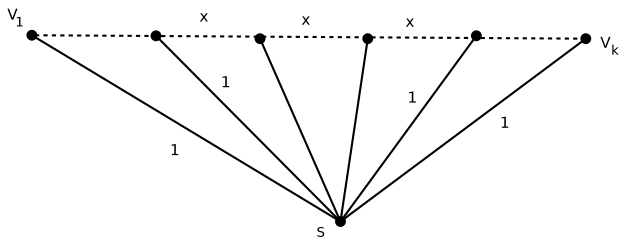}
\caption{The line circuit}
\label{fig:fincirc}
\end{figure}

Continuing our discussion of the circuit $S_{m}$, we increase the resistances connecting any two nodes adjacent to sink to the known upper bound of $|E(S)|-k$. Using the lemma \ref{lem:boundResVar}, we know that each of these increments decreases the value of $\pi_{v_{i}}(v_{j})$. Denote $x = |E(S)|-k$. Figure (\ref{fig:fincirc}) shows the circuit we have in the end. The value of $\pi_{v_{i}}(v_{j})$ obtained in this circuit will serve as a valid lower bound on the value we are seeking.

% OLD VERSION
% Continuing our discussion of the circuit $S_{m}$, we have already noted an upper bound on the resistances connecting any two nodes adjacent to sink and consecutive on the path $S_{m} - \{s\}$. Analyzing this circuit is somewhat intricate. We will use Lemma \ref{lem:boundResVar} to change the resistances such that the analysis becomes tractable. We increase the value of resistance connecting every pair of consecutive vertices in $S_{m} - \{s\}$, adjacent to $s$, to the known upper bound of $|E(S)|-k$. Using the lemma, we know that each of these increments decreases the value of $\pi_{v_{i}}(v_{j})$. Denote $x = |E(S)|-k$, then figure \ref{fig:fincirc} shows the circuit we have in the end. The value of $\pi_{v_{i}}(v_{j})$ obtained in this circuit will serve as a valid lower bound.

\begin{lem} Given the circuit $S_{m}$ as described above. For any pair of nodes $v$ and $w$ which are adjacent to the sink, the following upper bounds on the value of $\pi_{v}(w)^{-1}$ always hold.
\begin{eqnarray}
\max_{v,w}\{\pi_{w}(v)^{-1}\} = O((|E(S)|- k + 2)^{k-1})
\end{eqnarray}
\end{lem}

\textbf{Proof}: To keep the notation clean, we relabel the nodes in our circuit as follows. The node $v_{i}$ is $u_{1}$ and $v_{j}$ is $u_{k}$. All the nodes lying in between are indexed in order of occurrence on the path from $v_{i}$ to $v_{j}$. We apply unit potential at the site $u_{1}$ such that a potential of $\pi(u_{i})$ appears at node $u_{i}$, in particular $\pi(u_{1}) = 1$. Next, we scale the potential applied at $u_{1}$ so that unit potential appears at node $u_{k}$. Denoting the potential at node $u_{i}$ by $V_{i}$, Kirchhoff's equations of current conservation at any node $u_{i}$ is,

\begin{eqnarray}
 V_{i} & = & \frac{V_{i-1} + V_{i+1}}{x+2}\nonumber \quad  \forall 1 < i < k \\
 V_{k} & = & \frac{V_{k-1}}{x+1}\nonumber
\end{eqnarray}

which rearranges to give the recursive formulation,

\begin{eqnarray}
 V_{i} = (x+2)V_{i+1} - V_{i+2} \label{equ:recPot}
\end{eqnarray}

with the boundary condition

\begin{eqnarray}
 V_{k} & = & 1 \nonumber \\
 V_{k-1} & = & x+1 \nonumber
\end{eqnarray}

Consider the system, $V_{i}' = (x+2)V_{i+1}', V_{k}' = 1$. Then for each $i$, $V(i)' \geq V(i)$. Then $V_{1}' = (x+2)^{k-2}.(x+1)$. Therefore we have,

\begin{eqnarray}
\max_{v,w}\{\pi_{w}(v)^{-1}\} = O((|E(S)|- k + 2)^{k-1}) \nonumber
\end{eqnarray}

Note that the above solution is not far from the solution of the original set of equations.

\begin{eqnarray}
\left[ \begin{array}{c}
V_{1} \\
V_{2} \end{array}\right] =
\left[ \begin{array}{c c}
x+2 & -1 \\
1 & 0
\end{array}\right]^{k-2}
\left[ \begin{array}{c}
x+1 \\
1 \end{array}\right] \nonumber
\end{eqnarray}

The asymptotic eigenvalues of the matrix are $x+2$ and $0$ (for large $x$) and so the value of $V_{1}$ would be a linear combination of $(x+2)^{k-2}.(x+1)$ and some constant, which is asymptotically the same as our approximate solution. $\blacksquare$

We have already seen that,

\begin{eqnarray}
 tcl(S) = O(|E(S)|.\max_{v,w}\{\pi_{w}(v)^{-1}\}) \nonumber
\end{eqnarray}

So, for the case of a sandpile with $k$ connections to the sink, we have the Theorem \ref{thm:ksink}.

\noindent \textit{Remark}: We observed earlier in the introductory section that the line sandpiles have exponential transience classes. With slight amendment, the arguments used in proving the bounds stated above can be used to derive exponential lower bounds on the potential response in the line circuit. All one needs to do is replace the value of $x$ by $2$ and reduce the resistance of each connection to sink to half units. A completely combinatorial proof of the exponential nature of the transience class of line sandpiles appears in \cite{BGS}. 
\section{Equivalence of Triangular and Hexagonal Sandpile}

The definition of transience class describes it as the exact number of particles which surely induce a toppling everywhere in sandpile. In analogy with the question of time (or space) complexity of algorithms which asks for the maximum time taken by an algorithm, classifications exists on connected sets in $\mathbb{C}^{n}$ according to the maximum possible growth rates of continuous harmonic functions (the classical \textit{harnack's constant}) in terms of dimension and size of the set, upon graphs with respect to conductances, upon the speed of rumour spreading in graphs in terms of graph conductances \cite{CLP10}, \cite{CLP102}, upon graphs with respect to the growth rates harmonic functions itself (the harnack's constant in discrete setting), etc. Our goal is to impose a similar classification on sandpile families. In this section we will show that polynomial bounds on the transience class of one sandpile can be used to imply polynomial bounds on a related sandpile by considering the example of sandpiles based on honeycomb and triangular lattices.

An indexed family of sandpiles $\{ S_{n} \}$ is said to belong to the transience class $TCL(f(n))$ iff for all values of $n$

 \begin{eqnarray}
 tcl(S_{n}) = O(f(n)) \nonumber
 \end{eqnarray}

The transience classes $TCL(exp(n))$ and $TCL(poly(n))$ are defined in the usual manner. Our result on grid sandpiles establishes that $tcl(\chi_{n})$ belongs to $TCL(n^{7})$. We now introduce the notion of \textit{transience class equivalence}.

\begin{defin} We write $\{ A_{n} \} \sim_{tcl} \{ B_{n} \}$ if for any transience class $TCL(f(n))$, $\{ A_{n} \} \in TCL(f(n)) \Leftrightarrow \{ B_{n} \} \in TCL(f(n))$.\end{defin}

Two sandpile families $\{ A_{n} \}$ and $\{ B_{n} \}$ are transience class equivalent if they belong to the same transience classes. This formalises our intent to classify sandpiles into classes, where the number of particles needed for complete percolation is asymptotically equal, upto constant factors, for every sequence. This notion assumes importance in cases, when a sandpile-graph can be replaced by another sandpile-graph, equivalent in the above sense where transience class computations are easier to deal with. We will now show that a family of finite sandpiles based on honeycomb lattice, say  $\{ H_{n} \}$ belongs to $TCL(poly(n))$ iff the analogous family of finite sanpiles based on triangular lattice, say $\{ T_{n} \}$ belongs to $TCL(poly(n))$.

\begin{figure}[ht]
\centering
\includegraphics[width=.3\textwidth]{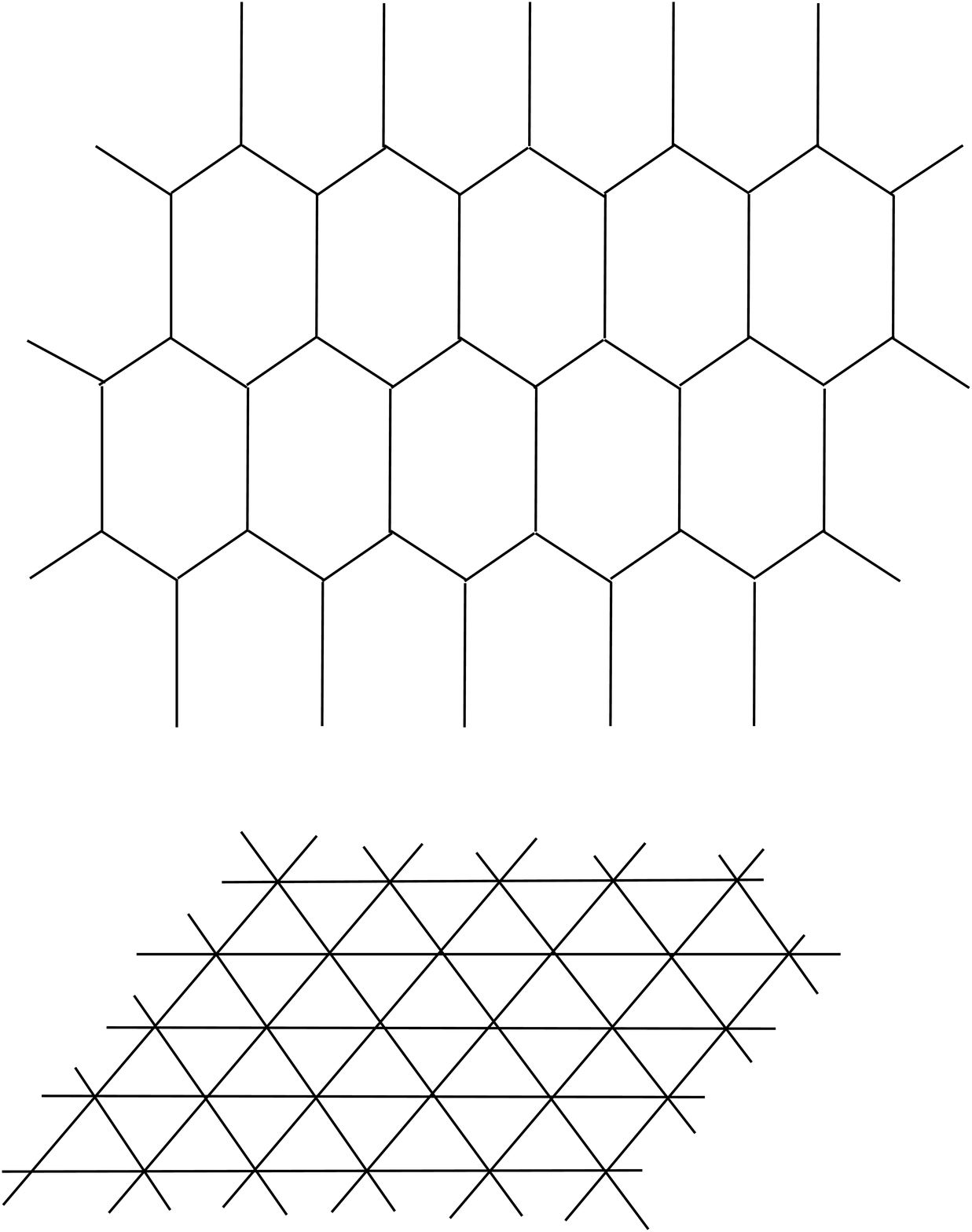}
\caption{Finite sections of honeycomb and triangular lattices}
\label{fig:honHex}
\end{figure}

In the simpler case of infinite (boundary-less) lattices, because of unbounded extension, it does not make sense to talk about the transience class. However, the sandpile impedence between any two sites is still well defined and is the right property to discuss. For planar lattices based on regular tessaletions of plane, these values can be estimated using simple particles conservation based combinatorial arguments. Given any planar lattice It is not too difficult to show that for any pair of vertices at a distance $n$ (shortest path length in the underlying graph), the value of $R_{S}(n)$ is $O(n^{2})$. The only property one needs is that the number of vertices in any region go up as the square of the radius of the region and some symmetry properties which are integral to regular tesselations. We now consider the case of finite honeycomb and triangular lattices with boundaries.

Figure \ref{fig:honHex} depicts finite sections of these lattices. The boundary edges, are connected to the sink node $s$ in both cases. Consider a sequence $\{ H_{n} \}$. We will construct the analogous sequence $\{ T_{n} \}$ whose membership in $TCL(poly(n))$ will imply membership of $\{ H_{n} \}$ as well. Let $H_{i}$ be any member. Consider the resistive circuit based on it, also referred to as $H_{i}$. This circuit will be transformed into an \textit{equivalent} circuit $T_{i}$. In the present context, equivalance will have a slightly more general meaning then in electric network theory.

\begin{defin}Two sandpile circuits $S_{1}$ and $S_{2}$ with the same boundary set $B (= \{v | v \sim s\})$ are said to be equivalent, if for any vertex $v \in B$, when unit potential is applied across $v$ and $s$, the potentials induced at all other vertices is identical in both cases. We denote network equivalence by $S_{1} \sim_{e} S_{2}$.\end{defin}

Following from lemmas \ref{prop:boundSand1} and \ref{prop:boundSand2}, one needs to consider only the vertices in the boundary set for obtaining bounds on $tcl$. Hence, when we say that the circuits $H_{i}$ is equivalent to $T_{i}$, the bounds on $tcl$ are identical. Since both the particles addition and last toppling nodes are on boundary, Lemmas \ref{lem:tclsil}, \ref{lem:tclsi} and the bounds on $\Gamma(.)$ ensure that if the $tcl$ is polynomial, the bounds obtained using Theorem \ref{thm:tcl} are also polynomial. So for sandpile sequences, the polynomial transience class is closed under equivalant reductions.

\begin{lem} Given $\{ A_{n} \}$ and $\{ B_{n} \}$, if $A_{i}\sim_{e} B_{i} $ for all values of $i$, then $\{ A_{n} \} \in TCL(poly(n)) \Leftrightarrow \{ B_{n} \} \in TCL(poly(n))$.\end{lem}

%For the special case when these are polynomial,

Before we start the reducing $H_{i}$, we will need the following result.

\begin{figure}[ht]
\centering
\includegraphics[width=.4\textwidth]{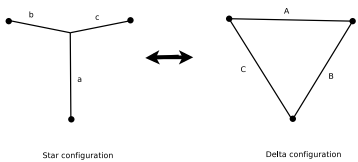}
\caption{The star-delta transformation}
\label{fig:starDelta}
\end{figure}

\begin{prop}(Star-Delta Transformation, \cite{Bol}) The configurations shown in figure \ref{fig:starDelta} are equivalent for
\begin{eqnarray}
 A = \frac{ab + bc + ca}{a} \nonumber
\end{eqnarray}
and likewise for the values of $B$ and $C$.
\end{prop}

\begin{figure}[ht]
\centering
\includegraphics[width=.3\textwidth]{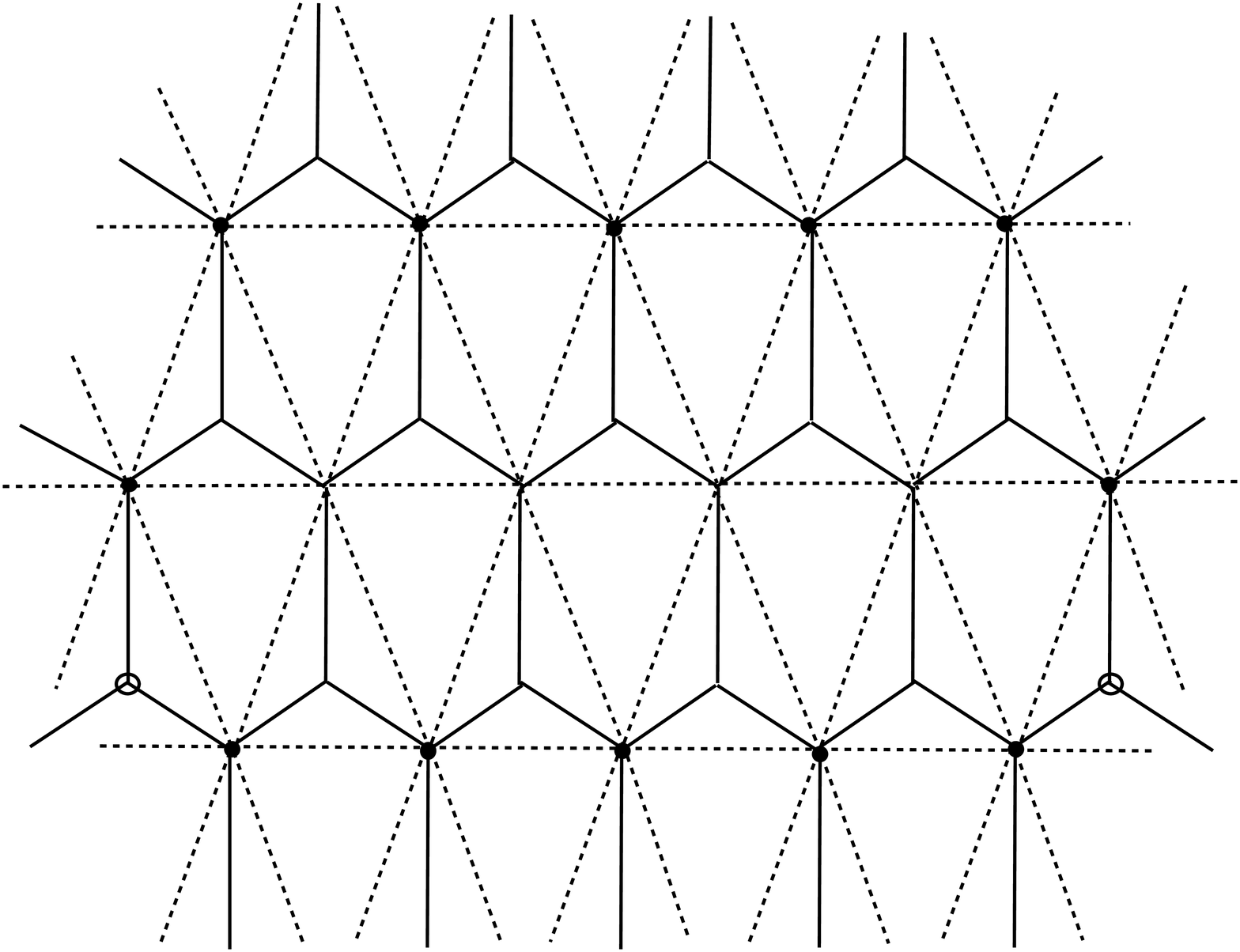}
\caption{A honeycomb based grid and its equivalent triangular lattice grid}
\label{fig:honHex-2}
\end{figure}

In the context of sandpiles, as long as the central node in the star configuration is not in the critical set (the concerned dense subset of boundary set), one can replace the configuration with the equivalent delta configuration without changing the potentials appearing on any boundary node when unit potential is applied at any boundary node. Consider the figure \ref{fig:honHex-2} which demonstrates a honeycomb lattice and its equivalent triangular lattice superimposed in dotted lines. The star configurations belong to the honeycomb lattice and are made up of unit resistances. The delta configurations (in dotted lines) constitute the delta configuration and each resistance has value $2$ units. Note that even if there exist boudary edges that do not belong to any complete star, they don't pose any essential problem as every unit resistance can be replaced with two $2$ unit resitors in parallel, as shown in the figure. The reduced triangular lattice we obtain is made up of $2$ unit resistors. Halving each resistor's value induces a constant factor change in the values of $\pi(.)$ and $\Gamma(.)$ functions over this circuit. The sandpile corresponding to this circuit is also denoted by $T_{i}$. We thus have a pair of sandpiles $T_{i}$ and $H_{i}$ such that membership of one in $TCL(poly(n))$ is equivalent to the membership of other.

% The polynomial bounds on transience class of grids derived by Babai and Gorodezky \cite{LB07} use the $D_{4}$ symmtery of grids in an essential way. The presence of $D_{6}$ symmetry in triangular lattices allow the exact same proof methodology to carry over from their work and hence imply polynomial bounds on the transience class of suitably shaped finite regions on triangular lattices. Likewise with the honeycomb lattice whose symmetry group is $C_{3}$ (cyclic with order $3$). The only allowance one needs to make in the discussions already made in \cite{LB07}, is to change the shapes of the travelling diamonds and boxes. In case of triangular girds, these are hexagonal while in the case of honeycomb, they are triangular.

The reductions we display above, prove important in the cases when only one of the members of an equivalent pair has the necessary symmetries to deduce polynomial bounds.

% \noindent \textit{Remark}: Intuitively it is not at all plausible for the symmetry group itself to play any crucial role in determining diffusion speeds. In a related work, we return to this theme and show that the
%It is well known that the square lattice, triangular lattice and honeycomb lattice are the only regular planar tessalations. The square lattice is its own dual while the honeycomb and triangular lattice are duals of each other. Almost tautologically, square lattice is transience-class equivalent to its geometric dual, such a statement about the other two tilings is not so straightforward. We will now demonstrate the equivalence of the sandpiles on these two tilings. 
\section{Future work and Open problems}

The main open question is that of tightening the bounds on $tcl(\texttt{GRID}_{n})$. As noted in the remark at the end of subsection \ref{sec:gridBound}, one can expect  substantial improvements only in the estimation of $\max_{v,w}\pi_{w}(v)^{-1}$. We believe the approach using Lemma \ref{lem:lapCurr} is the most promising avenue.

The general form of the eigenvalues and eigenvectors of a grid are well known. Using these and the results in Lemma \ref{lem:lapCurr}, one can approximate (up to constant factors) the corner to corner potential correlation using the following function.

\begin{eqnarray}
V =  \frac{1}{n^{2}}\sum_{\substack{0 \leq a < b \leq n-1 \\ a \neq b}}\frac{ (-1)^{a+b+1}\sin^{2}{\frac{(a-b)\pi}{2n}}\sin^{2}{\frac{(a+b)\pi}{2n}}\cos^{2}{\frac{a\pi}{2n}}\cos^{2}{\frac{b\pi}{2n}} }{4 - 2\cos (\frac{a\pi}{n}) - 2 \cos (\frac{b\pi}{n})}\nonumber
\end{eqnarray}

After spending considerable time on trying to resolve the absolute size of this expression, which started as a seemingly harmless looking question and subsequently led to a formulation of sorts on the general size estimation problem of alternating sums based on uniformly continuous functions, the authors must admit their inability in resolving this rather technical problem and invite the interested reader from the theory community to take it up from this expression. For purposes of restricting the manuscript size, we have not described the complete algebra leading to this expression as well as the details of our estimation procedure, these are available on personal request.

The second open question is showing that the opposite corner sites are indeed the worst pair for single site particle addition strategies. The more general question of finding the analogous pair in general graphs is also interesting. We conjecture the following implicit characterization of such a pair.

\begin{conj} For a given sandpile $S$, if the site $u$ allows one to attain the worst case bounds for single site particle addition strategies, then in the corresponding circuit, there exists a boundary site $w$ such that $\pi_{w}(v) = min_{p,q}\pi_{p}(q)$.
\end{conj}

This hints at another of the many ways in which sandpile are similar to electric networks.We have already demonstrated that single site particle addition strategies are not enough to attain the transience class. We believe however that the following weaker conjecture holds.

\begin{conj}\label{conj:tclSing} Assume that for sandpile $\chi$, the induced sub-graph on the set of ordinary vertices is connected. Then the transience class of $\chi$ (the largest weight of any transient configuration) is bounded from above by the sum of height of the tallest transient stack of grains placed on a single site and the size of graph $\chi$.\end{conj}

\noindent \textit{Remark:} One can show that the transience class of a sandpile is bounded from below, up to constant factor, by the graph size. This follows from the fact that when every site has toppled, no two adjacent sites can both have zero particles. Which implies that for every edge, at least one particle is on board. So $tcl(\chi) = \Omega(|\chi|)$. In the light of this observation, conjecture \ref{conj:tclSing} means that the estimates derived using single site particle addition strategies are constant additive factor approximations of the actual transience class. Also note that the above conjecture is stronger then the question raised by Sunic, mentioned in \cite{BT05}, that local (i.e. single site particles addition type) transience classes are bounded by $0.5$-factor approximations of the transience class, as we have an additive error term compared to the previous multiplicative one.

% The general form of the eigenvalues and eigenvectors of a grid indexed by ordered pair $(i,j)$ with $0 \leq i,j \leq n-1$ are well known.
% \begin{eqnarray}
% \lambda_{i,j} = 4 - 2\cos (\frac{i\pi}{n}) - 2 \cos (\frac{j\pi}{n})
% \end{eqnarray}
% and the corresponding eigenvectors are,
% \begin{eqnarray}
% \psi_{0,0} = \frac{1}{n} [1 1 1 \ldots 1]
% \end{eqnarray}
% \begin{eqnarray} \label{equ:eigenvec}
% \psi_{(i,j),(x,y)} = \frac{2}{n}\cos{\frac{(x+1/2)i\pi}{n}}\cos{\frac{(y+1/2)j\pi}{n}}
% \end{eqnarray}
% We make a minor change in the placement of the current source. Instead of connecting it through the vertex $(0,0)$, we connect the source at node $(0,1)$ and sink at $(1,0)$.
% Putting these values in the equation (\ref{equ:currSize}) and simplifying, we obtain the following expression.
% \begin{eqnarray}
% i_{e} =  |\frac{128}{n^{2}}\sum_{\substack{0 \leq a < b \leq n-1 \\ a \neq b}}\frac{ (-1)^{a+b+1}\sin^{2}{\frac{(a-b)\pi}{2n}}\sin^{2}{\frac{(a+b)\pi}{2n}}\cos^{2}{\frac{a\pi}{2n}}\cos^{2}{\frac{b\pi}{2n}} }{4 - 2\cos (\frac{a\pi}{n}) - 2 \cos (\frac{b\pi}{n})}| \nonumber
% \end{eqnarray}

\section{Acknowledgements}
The authors thank Laszlo Babai for his first review of this manuscript. We extend our heartfelt gratitude to Gopal Srinivasan, Milind Sohoni, Bharat Adsul, Jugal Garg, Ruta Mehta and Nutan Limaye for the many helpful discussions, suggestions, comments and criticisms throughout the course of this study.
A very special thanks to Zahra Jafargholi for her critical reading of this paper's preliminary version and helping us catch some basic errors.

\end{document}